\documentclass[]{aa}  

\usepackage{natbib}
\bibpunct{(}{)}{;}{a}{}{,}
\usepackage{graphicx}
\usepackage{revsymb}	
\usepackage{amsmath}	
\usepackage[usenames]{color}	
\usepackage{txfonts}
\usepackage{amssymb}
\usepackage{color}
\usepackage{multirow}
\usepackage{calrsfs}
\usepackage{geometry} 
\usepackage[parfill]{parskip} 
\usepackage{amssymb}
\usepackage{slantsc}
\usepackage{array}
\usepackage{url}
\usepackage{amsfonts}
\usepackage[colorlinks=true,linkcolor=black, urlcolor=black, citecolor=black]{hyperref}
\usepackage{hyperref} 
\usepackage{sidecap}
\usepackage{graphicx}
\usepackage{xcolor}
\usepackage{nicefrac}
\usepackage{subfigure}
\usepackage{epsfig}
\colorlet{rouge}{red!70!darkgray}
\titlerunning{Internal rotation of Kepler 56}
\usepackage{dutchcal}

\begin{document}

   \title{Asteroseismology of evolved stars to constrain the internal transport of angular momentum}

\subtitle{IV. Internal rotation of Kepler 56 from an MCMC analysis of the rotational splittings}

\author{L. Fellay\inst{1} \and G. Buldgen\inst{1} \and P. Eggenberger\inst{1}  \and S. Khan\inst{2} \and S.J.A.J. Salmon\inst{1} \and A. Miglio\inst{2,3,4}\and J. Montalbán\inst{2} }

\institute{Observatoire de Genève, Université de Genève, Chemin Pegasi 51, CH$-$1290 Sauverny, Suisse \and  School of Physics and Astronomy, University of Birmingham, Birmingham B15 2TT, United Kingdom \and Dipartimento di Fisica e Astronomia, Università di Bologna, Via Gobetti $93/2$, $I-40129$ Bologna, Italy \and INAF $--$ Astrophysics and Space Science Observatory Bologna, Via Gobetti $93/3$, $I-40129$ Bologna, Italy}
\date{November, 2020}
\abstract{The observations of global stellar oscillations of post main-sequence stars by space-based photometry missions allowed to directly determine their internal rotation. These constraints have pointed towards the existence of angular momentum transport processes unaccounted for in theoretical models. Constraining the properties of their internal rotation thus appears as the golden path to determine the physical nature of these missing dynamical processes.}
{We wish to determine the robustness of a new approach to study the internal rotation of post main-sequence stars, using parametric rotation profiles coupled to a global optimization technique.}
{We test our methodology on Kepler 56, a red giant observed by the \textit{Kepler} mission. First, we carry out an extensive modelling of the star using global and local minimizations techniques, and seismic inversions. Then, using our best model, we study in details its internal rotation profile, we adopted a Bayesian approach to constrain stellar parametric predetermined rotation profiles using a Monte Carlo Markov Chain analysis of the rotational splittings of mixed modes. }
{Our Monte Carlo Markov Chain analysis of the rotational splittings allows to determine the core and envelope rotation of Kepler 56 as well as give hints about the location of the transition between the slowly rotating envelope and the fast rotating core. We are able to discard a rigid rotation profile in the radiative regions followed by a power-law in the convective zone and show that the data favours a transition located in the radiative region, as predicted by processes originating from a turbulent nature such as for example magnetic instabilities. }
{Our new approach to study the internal rotation of red giants constitutes a viable option to analyse \textit{Kepler} targets and allows us to put stringent constraints on the properties of the missing angular momentum transport process acting in post main-sequence stars. Our analysis of Kepler 56 indicates that turbulent processes whose transport efficiency is reduced by chemical gradients are favoured, while large scale fossil magnetic fields are disfavoured as a solution to the missing angular momentum transport.}
\maketitle

\section{Introduction}
Thanks to the advent of long space-based photometric surveys such as CoRoT \citep{CoRoT2009}, \textit{Kepler} \citep{KEPLER2010} and TESS \citep{TESS2014, TESS2015}, asteroseismology has become the most efficient technique to test the theory of stellar structure and evolution. In that respect, it serves as the golden path to analyze the dynamical processes acting in stellar interiors. One of the key results of the space-based photometry revolution is the detection of the so-called mixed modes in post main-sequence stars, that behave as acoustic modes in the outer layers and gravity modes in the deep layers. Their dual nature enabled the determination of the internal rotation of subgiant and red giant stars \citep{Deheuvels2012, Deheuvels2014, Deheuvels2015, Deheuvels2020, Beck2012, Mosser2012, DiMauro2016, DiMauro2018, Gehan2018} that pointed towards a very efficient angular momentum (AM) transport process absent from theoretical stellar models \citep{Eggenberger2012,Eggenberger2017, Eggenberger2019I, Ceillier2013, Marques2013, Cantiello2014, Spada2016}.

Various transport mechanism candidates have been suggested to reproduce asteroseismic data, from internal gravity waves \citep{Pincon2017}, mixed modes \citep{Belkacem2015I,Belkacem2015II}, magnetic instabilities \citep{Spada2016, Fuller2019, Eggenberger2019II, denHartogh2020} and internal fossil magnetic fields \citep{Kissin2015, Takahashi2020}. These processes are still largely investigated by the community with the use of stellar evolution codes including a proper treatment of AM transport during evolution. The issue is quite tedious as even non-standard processes invoked to reproduce the solar rotation profile deduced from helioseismic measurements \citep[e.g.][]{Eggenberger2005, Eggenberger2019} prove not efficient enough in later stages \citep{Cantiello2014,denHartogh2019}. However, some of these studies have only focussed at reproducing the core rotation of red giants, while tighter constraints can be drawn from the degree of differential rotation in the star \citep{Deheuvels2014} and the localization of the transition between the slow rotating upper layers and the faster rotating core \citep{DiMauro2018}. The signature of the location of this transition is deeply linked to the pulsation properties of the star and the extent of the gravity and pressure cavities. Determining precisely the location of the transition in rotation and discriminating between various forms of the internal rotation profile of these red giants is crucial to constrain the physical nature of the missing transport mechanism.

In this study, we investigate the potential of parametric rotation profiles, coupled to a Monte Carlo Marko Chain (MCMC) algorithm to determine the properties of the internal rotation profile of the red giant  star Kepler-56 (KIC6448890) studied by \cite{Huber2013}. In the case of Kepler-56, we are seeing a good example of strong coupling between the two pulsation cavities. This case has been recently studied by \cite{Takata2016b, Takata2016a}, who derived the asymptotic properties of mixed modes in this regime. Previous studies referring to this case can be found in \citet{Mosser2012,Goupil2013}. The case of strong coupling is characterised by the fact that one p-mode is coupled to a whole range of g-modes of neighbouring frequencies.

As a good depiction of the internal structure is required for a detailed analysis of the internal rotation, we start by carrying out a detailed structural modelling of the star, presented in Sect. \ref{Stellar_model}. In Sect. \ref{MCMC_analysis}, we describe the MCMC analysis and demonstrate its robustness on synthetic data, while the detailed analysis of Kepler-56 is carried out in Sect. \ref{Kepler56} and the consequences of our results for the physical nature of the missing transport mechanism in stellar interiors are discussed in Sect. \ref{subsec_AM}. The conclusion as well as the potential of our approach for other targets is discussed in Sect.~\ref{Conclusion}.

\section{Stellar models and properties}\label{Stellar_model}
To be able to study the internal rotation of Kepler 56, we first need to determine a reliable model of its internal structure, so that the first order perturbation analysis of the detected rotational splittings is meaningful. To that end, we combine global and local minimization techniques using seismic and non-seismic constraints, as well as mean-density inversions, to obtain a robust model of our target.

\subsection{Observational constraints}
The star was modelled based on asteroseismic and classical constraints as in \cite{Huber2013}. We summarize in Table~\ref{table_classical_constraints} the global seismic indexes as well as the classical constraints for Kepler 56, while a table with the full seismic data adopted is given in Appendix \ref{table_data_sismo}.  

\begin{table}[h]
\centering
\caption{Classical observational constraints obtained by spectroscopy on the top panel. Global seismic parameters are presented on the bottom panel \citep{Huber2013}.}\label{table_classical_constraints}
\begin{tabular}{ll}
\hline\hline
\multicolumn{2}{l}{\multirow{2}{*}{Observational constraints}} \\
\multicolumn{2}{l}{}                                       \\ \hline
Metallicity [Fe/H]            & $0.20 \pm 0.16$      \\
Effective Temperature {[}$K${]}     & $4840 \pm 97$        \\
Luminosity {[}$L_\odot${]}          & $8.602 \pm 0.363$    \\ \hline
\multicolumn{2}{l}{Large separation $\Delta\nu=17.4\pm0.1\ \mu \rm{Hz} $}             \\
\multicolumn{2}{l}{Frequency of max. power $\nu_{max}=244.3\pm 1.3\ \mu \rm{Hz} $} 
\\ \hline
\end{tabular}
\end{table}

The luminosity presented in Table \ref{table_classical_constraints} was computed from the following formula 
\begin{equation}
\log\left(\frac{L}{L_\odot}\right) = -0.4\left(m_\lambda + BC_\lambda -5\log d + 5 - A_\lambda -M_{\mathrm{bol},\odot}\right) \, ,
\end{equation} 
where $m_\lambda$, $BC_\lambda$, and $A_\lambda$ are the magnitude, bolometric correction, and extinction in a given band $\lambda$. We use the 2MASS $K$-band magnitude properties. The bolometric correction is estimated using the code written by \citet{Casagrande2014,Casagrande2018}, and the extinction is inferred with the \citet{Green2018} dust map. A value of $M_{\mathrm{bol},\odot} = 4.75$ is adopted for the solar bolometric magnitude. Using Gaia DR3 \citep{Gaia2020}, we obtain a luminosity value of $L=8.602 \pm 0.363\ L_{\odot}$. This value is however preliminary, as we could not take into account the systematic errors of the Gaia DR3 catalog, as well as for a correction of the zero-point parallax offset.

\subsection{Stellar modelling}
The stellar evolutionary models used in this study and the corresponding oscillation frequencies were computed with the Code Liégeois d'Evolution Stellaire \citep[CLES,][]{Scuflaire2008a} and the Liège OScillation Code \citep[LOSC,][]{Scuflaire2008b}. 
\\Our modelling is divided into three distinct phases:  first, we carry out the modelling using classical constraints and the individual radial frequencies using the AIMS software \citep{AIMS2016, AIMS2019}. Second, this initial step is used to determine an inverted mean density value, that is then used in the third modelling step using a Levenberg-Marquardt minimization technique \citep{Levenberg1978} and aiming at reproducing the whole oscillation spectrum of Kepler 56, computing the evolutionary models on the fly rather than using a predetermined grid of models coupled with an interpolation method, as in AIMS.

This last modelling step is a requirement to make sure that the properties of the mixed modes are well reproduced and to avoid non-linearities in the variational formulations used for the analysis of the rotational splittings of these modes. Indeed, looking at Eq.~\ref{eq_splitting}, we can see that the integral relation between the rotational splittings and the rotation profile will also depend on the kernel function. In practice, this function depends also on the properties of the eigenfunction. In the case of mixed modes, a small shift in frequency can be associated with a large variation of the eigenfunction, and lead to biased estimates of the internal rotation (see discussion in Sect.~\ref{subsec_splitting}).

The grid properties for the AIMS modelling are summarized in Table~\ref{table_model_param}. This grid has been tailored to encompass the results for the mass and radius given by the seismic scaling relations and the surface metallicity given in Table~\ref{table_classical_constraints}. We used the AGSS09 \citep{Asplund2009} solar abundances with a corrected abundance of Ne/O denoted AGSS09Ne \citep{Landi2015, Young2018} for AIMS, while the Levenberg-Marquardt modelling is using the classical AGSS09 abundances\footnote{Given the large uncertainties on the spectroscopic constraints for Kepler 56, the small change in the solar abundance tables has no impact on the final results.}. For both modelling steps, the models use the FreeEOS equation of state \citep{Irwin2012}, the OPAL opacities \citep{Iglesias1996}, the $T(\tau)$ relation from Model-C of \citet{Atmosphere1981} for the atmosphere, the Mixing Length Theory of convection implemented as in \citet[][]{Cox1968} and the nuclear reaction rates of \cite{Reaction2011}. The constraints and free parameters of both the AIMS and Levenberg-Marquardt modelling steps are summarized in Table~\ref{table_model_param}. In this table, $X_0$ denotes the initial hydrogen mass fraction, $Z_0$ the initial metal mass fraction, [Fe/H] the observed metallicity, $L$ the observed luminosity and $M$ the mass. The convection is controlled by two parameters, the classical mixing-length parameter $\alpha_{MLT}$ and $\alpha_{\rm{over}}$ characterising the length of the core overshooting region with the same formalism as the MLT. The temperature gradient in the overshooting region is assumed adiabatic in both modelling steps. The mixing-length parameter was kept at a solar calibrated value for all tracks of the AIMS grid. While fitting the individual radial modes, we took into account the impact of surface effects by using the two-term surface correction of \cite{Ball2014} in AIMS. Uniform priors on the width of the grids were used for the MCMC modelling with AIMS. We provide the grid properties in table \ref{table_model_param}.

\begin{table}[h]
\centering
\caption{Summary of the constraints and free parameters used for both AIMS and the Levenberg-Marquardt modelling steps. Here $\nu_{cross}$ denote the frequency of the avoided crossing of lowest frequency \citep{Deheuvels2011}. $a_3$ and $a_{-1}$ denotes the surface correction coefficient defined in \cite{Ball2014}. $\mathcal{U}$ denotes uniform priors.   }\label{table_model_param}
\begin{tabular}{clll}
\hline\hline
\multirow{2}{*}{}                                    & \multirow{2}{*}{AIMS} &\multirow{2}{*}{AIMS's priors}& \multirow{2}{*}{Levenberg} \\
                                                     &                       &      &                      \\ \hline
\multirow{6}{*}{Constraints}                         & [Fe/H]        &         & $L$                   \\
                                                     & $\nu_{max}$        &   & $\overline{\rho}$              \\
                                                     & $5\ \nu_{\ell=0}$   &           & $T_{\rm{eff}}$                  \\
                                                     & $T_{\rm{eff}}$        &     & [Fe/H]                     \\
                                                     &               &        & First $\nu_{\ell=0}$       \\
                                                     &                 &      & $\nu_{cross}$              \\ \hline
\multicolumn{1}{l}{\multirow{6}{*}{Free parameters}} &  $M$  & $\mathcal{U}[1.00-2.22]\ M_\odot$            & $M$                   \\
\multicolumn{1}{l}{}                                 & Age        &  $\mathcal{U}[0-14]\ Gyr$         & Age                        \\
\multicolumn{1}{l}{}                                 & $X_0$    &      $\mathcal{U}[0.68-0.72]$       & $X_0$                      \\
\multicolumn{1}{l}{}                                 & $Z_0$      &   $\mathcal{U}[0.010-0.034]$        & $Z_0$                      \\
\multicolumn{1}{l}{}                                 & $a_3$      &      no priors     & $\alpha_{MLT}$             \\
\multicolumn{1}{l}{}                                 & $a_{-1}$     &    no priors        & $\alpha_{over}$            \\ 
\hline
\end{tabular}
\end{table}

The inversion procedure of the mean density is then used to derive a model-independent mass interval that is used as a safeguard for the following modelling steps. In practice, as this inversion is only computed using radial modes following the approach of \cite{Buldgen2019}, we do not need to worry about potential non-linearities at this stage. The results for the inverted mean density and the model-independent mass interval derived from Gaia and spectroscopic data is given in Table~\ref{table_stellar_properties}. The model-independent mass interval was derived by combining the inverted mean density to the stellar radius obtained though the expression of the luminosity of a black body using the Gaia luminosity and spectroscopic stellar effective temperature.
 As we can see, the modelling results lie well within this mass interval, confirming the robustness of our procedure.

In the last modelling step, we used the constraints and free parameters listed in the right column of Table~\ref{table_model_param}. The modelling follows a similar approach as the one introduced by \cite{Deheuvels2011}, but having here the same number of free parameters as constraints. Before attempting to reproduce the dipolar mixed modes, a run of the Levenberg-Marquardt minimization aiming at reproducing the so-called individual frequency ratios was carried out, serving as an intermediate step. These ratios are sensitive to the stellar mass for a given mean density \citep{Montalban2010}, and are defined as
\begin{align}
r_{02}=\frac{\nu_{n,0}-\nu_{n-1,2}}{\langle\Delta \nu\rangle},
\end{align}
where $\langle\Delta \nu\rangle$ is the average large frequency separation deduced from a linear fit of the radial modes (using the definitions in \citet{Reese2012den}). Once a good agreement is obtained for the frequency ratios, we turn our attention to reproducing the g-dominated dipolar frequency of the avoided crossing of lowest frequency. Namely here we ensure that the g-dominated mode of frequency $192.402$ $\mu \rm{Hz}$, denoted  $\nu_{cross}$  in Tab.~\ref{table_model_param}, is well-reproduced by the model. The final verification step consists in the analysis of the Echelle diagram for the model, which ensures that we reproduce with a sufficiently good accuracy the global oscillation pattern. 

\begin{figure}[h]
\centering
\includegraphics[width=\hsize]{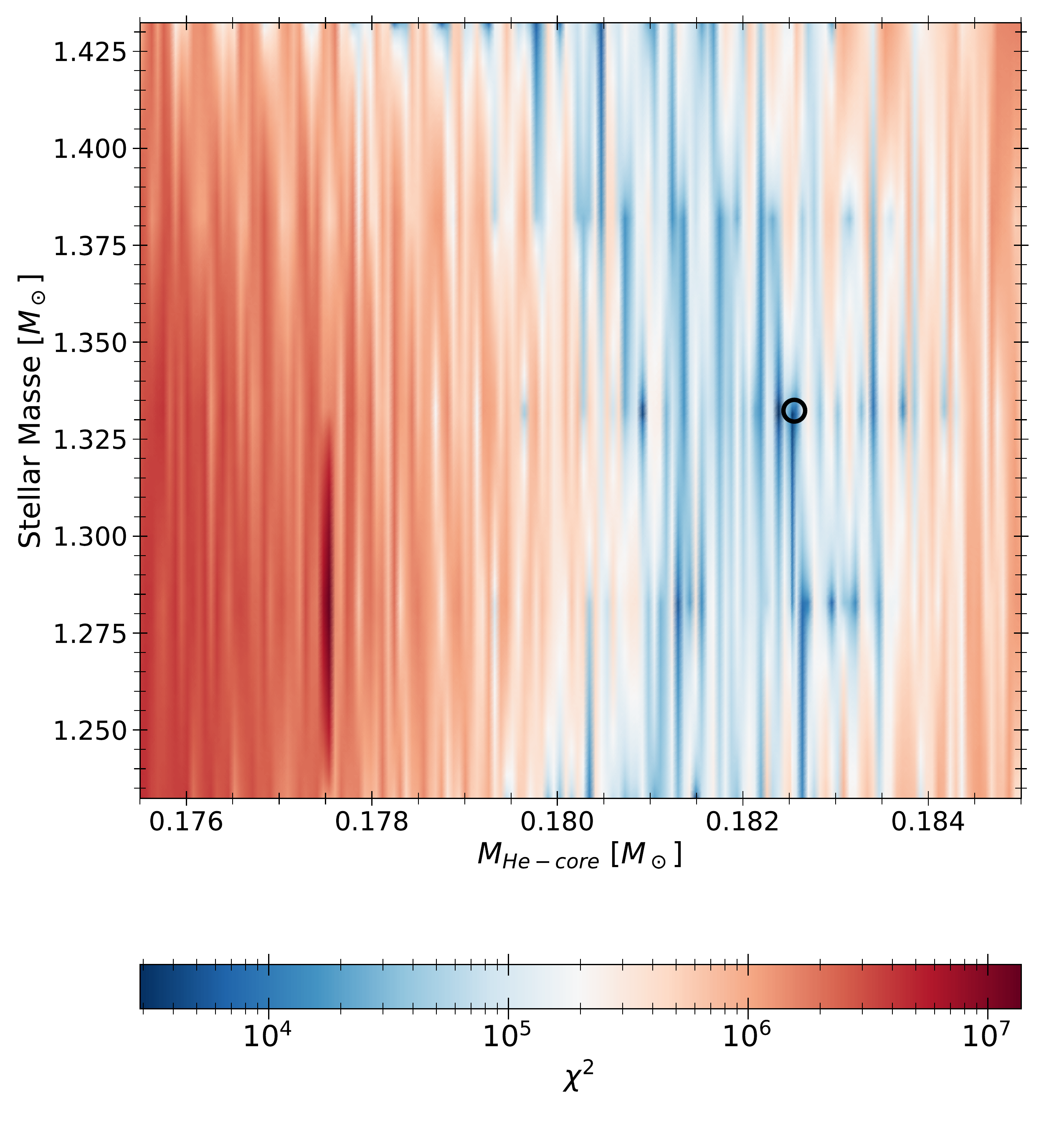}
\caption{Mapping of the parameter space around the model carried out with the Levenberg-Marquardt technique. The black circle represents the position of the best fitting model found with the Levenberg-Marquardt minimization technique.}
\label{fig._chi2_map}
\end{figure}

The fit was carried out starting from various mass values to better explore the parameter space and we determined the uncertainties in our modelling from an analysis of a $\chi^{2}$ map where we considered the constraints given in column 2 of Tab.~\ref{table_model_param} and the individual frequency ratios $r_{02}$ in the evaluation of the $\chi^{2}$.
The map was carried out using a grid of $5$ different masses, $3$ metallicities, $3$ initial hydrogen and $3$ $\alpha_{MLT}$, all centered around the value found for our best fitting model. The models were limited in age by the Gaia luminosity given in Tab.~\ref{table_classical_constraints}. A proxy for the age, the mass of the helium core, was used to avoid problems linked to the extreme age differences between tracks caused by the initial hydrogen, metallicity or mass. Finally an interpolation in mass and helium core mass was made to produce the $\chi^{2}$ map shown in Fig.~\ref{fig._chi2_map}.

A valley of minima can clearly be seen in Fig. \ref{fig._chi2_map} with our model as the overall minimum. A second clear minimum can be seen in the models with masses of $1.38\ M_\odot$, but the $\chi^{2}$ value for our optimal model is two orders of magnitude lower and appears as a clear minimum in the valley observed in the parameter space. This study confirms the robustness of our stellar model and of the determined stellar parameters.

\subsection{Stellar properties}

\begin{figure}[h]
\centering
\includegraphics[width=\hsize]{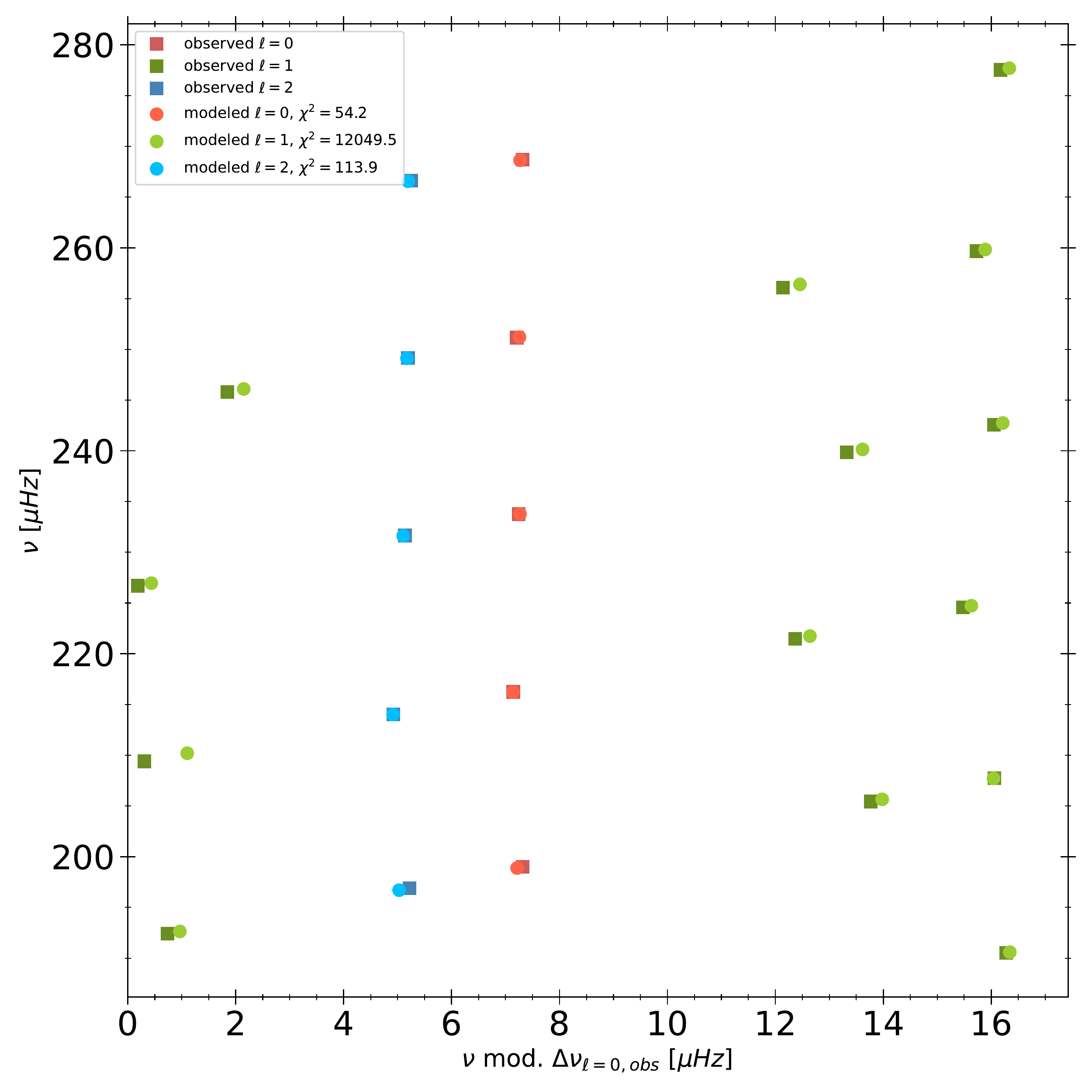}
\caption{Echelle diagram of Kepler 56 comparing observed oscillation frequencies (listed in Table~\ref{table_data_sismo}) represented as squared points and theoretical oscillation frequencies represented as circle points. The difference between observed and modelled data is quantified, in the legend, as a $\chi^2$ for each spherical order $\ell$. Finally, the error bars on the observed frequencies are included in the data points and not clearly distinguishable here.}
\label{fig._echelle_diagram}
\end{figure}

\begin{figure}[h]
\centering
\includegraphics[width=\hsize]{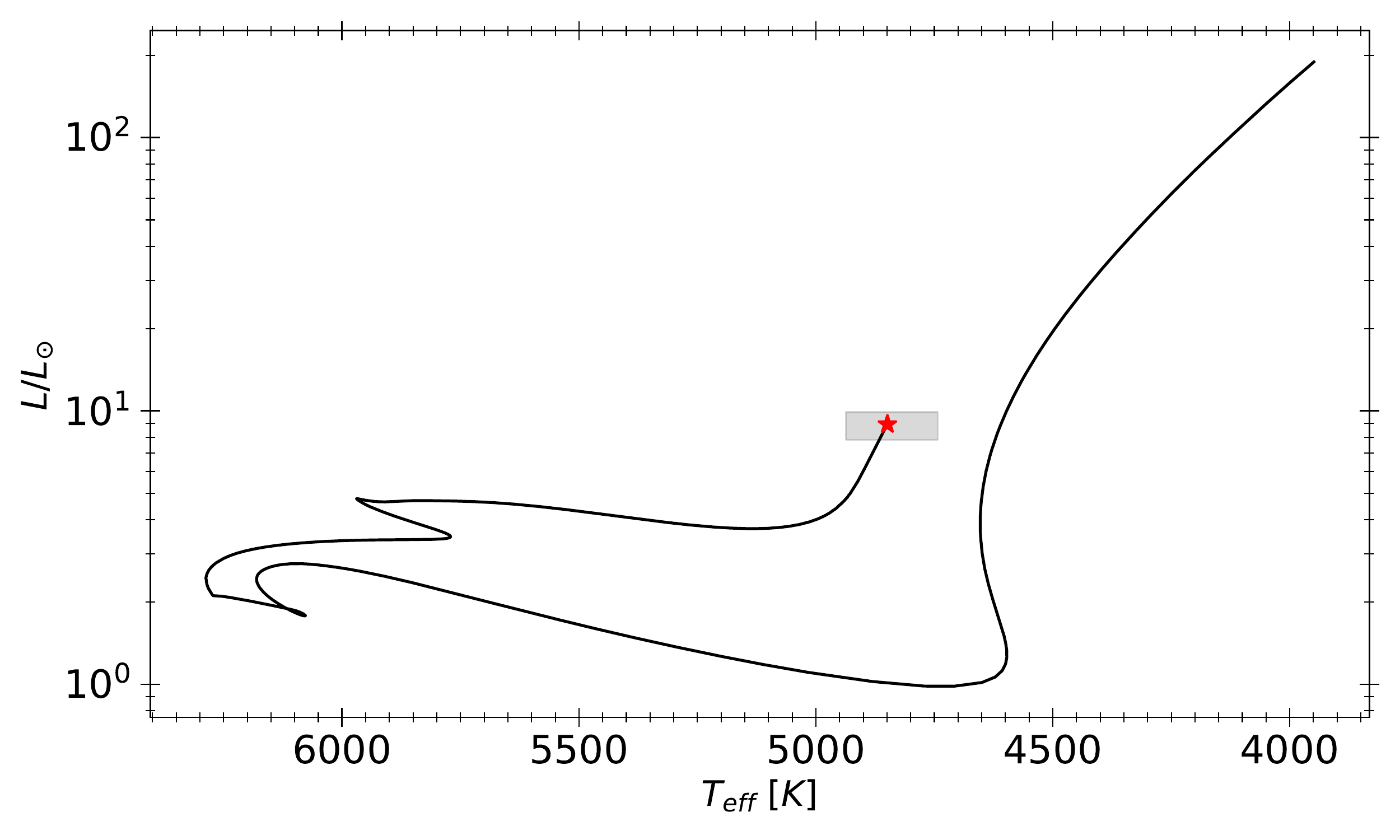}
\caption{Hertzsprung–Russell diagram of the evolution of Kepler 56 from the pre-MS to the modelled actual state of the star. The grey area corresponds to the constraints on the effective temperature and luminosity listed in Table~\ref{table_classical_constraints}. }
\label{fig._hr_diagram}
\end{figure}

Ultimately, the final solution of our structural modelling is given in Tab.~\ref{table_stellar_properties}, while the Hertzsprung–Russell diagram of the star modelled with the Levenberg-Marquardt minimization technique is presented in Fig.~\ref{fig._hr_diagram}. The agreement obtained in terms of pulsation frequencies is illustrated by the Echelle diagram shown in Fig.~\ref{fig._echelle_diagram} for the whole pulsation spectrum. 

\begin{table*}[h]
\centering
\caption{Non-seismic stellar properties of Kepler 56 obtained in the different modelling steps.}\label{table_stellar_properties}
\begin{tabular}{llll}
\hline\hline
\multirow{2}{*}{Stellar parameters} & \multirow{2}{*}{AIMS} & \multirow{2}{*}{Density Inversion+ Gaia + spectroscopy} & \multirow{2}{*}{Levenberg} \\
                                    &                       &                                    &                        \\ \hline
Mass {[}$M_\odot${]}                & $1.286 \pm 0.011$          & $1.342 \pm 0.181$                       & $1.332 $           \\
Radius {[}$R_\odot${]}              & $4.179 \pm 0.132$          &$4.244 \pm 0.189$                      & $4.235  $           \\
Mean Density {[}$g/cm^3${]}         & $ 0.0248\pm 0.0001$        & $ 0.0247 \pm 0.0005$                     & $ 0.0247 \ $         \\
Metallicity $Z_0/X_0$           & $0.0251 \pm 0.013$         & \textunderscore                      & $0.0314  $          \\
Effective temperature {[}$K${]}     & $4973 \pm 14$           &$4840 \pm 97$                         & $4849  $            \\
Age {[}$Gyr${]}                     & $3.917 \pm 0.157$          & \textunderscore                       & $5.605  $           \\
Luminosity {[}$L_\odot${]}          & $ 9.589 \pm 0.129$         &$8.602 \pm 0.364$                       & $ 8.934  $          \\
$\alpha_{MLT}$                      & $2.031$             &\textunderscore                         & $2.125 $              \\
$\alpha_{\rm{over}}$                     & $0 $            &\textunderscore                         & $0.1829 $             \\ \hline
\end{tabular}
\end{table*}

The classical stellar properties obtained with the Levenberg-Marquardt method are within a $1-\sigma$ difference with respect to the ones presented by \cite{Huber2013}.  A direct comparison of the models of both studies using a $\chi^2$ function is irrelevant here since different constraints were used in the fit. However, we can compare the agreement in terms of individual frequencies for both the model computed with AIMS, fitting only the radial modes and the one computed with the Levenberg-Marquardt technique, taking into account non-radial oscillations. Unsurprisingly, we then see that the latter has a much better $\chi^2$ than the former, namely by an order of magnitude for the $\ell=2$ modes and by a factor $3$ for the $\ell=1$ modes.

We mention that the Echelle diagram shown in Fig.~\ref{fig._echelle_diagram} includes the empirical surface corrections of \cite{Ball2014} for the $\ell=0$ and the p-dominated $\ell=1$ modes. We mention that these corrections were calibrated from acoustic oscillations of main-sequence stars.
By fitting only the lowest frequency modes, we ensured a limited impact of the surface effect correction on the final results, while avoiding additional parameters for the modelling. Comparing Fig. S8 of \citet{Huber2013}, we can see that our solution is as good as theirs, given that they included surface corrections in their direct modelling of the individual frequencies. Overall, this model shows a good agreement in both individual frequencies and frequency ratios $r_{02}$.

We also see that our optimal mass value agrees well with the model-independent interval defined using the mean density inversion. Our global parameters agree within one sigma with the results of the detailed modelling of \citet{Huber2013}, despite the use of the different stellar evolution codes \citep{ATON2008} and reference solar abundances \citep{Grevesse1998}. This lends us further confidence in the robustness of our results and the agreement with the individual frequencies is suitable enough to carry out the analysis of the rotational properties of the mixed modes.

\section{MCMC analysis}\label{MCMC_analysis}
In this section, we detail our procedure regarding the MCMC analysis of the rotation profile of post main-sequence stars. First, we introduce in Sect. \ref{subsec_splitting} the theoretical brackgound of our approach as well as the basis equations applied in the MCMC analysis. In Sect. \ref{subsec_MCMC_method}, we detail our methodology for our MCMC analysis. Finally, in Sect. \ref{subsec_MCMC_confirmation} we prove the robustness of our method by testing the recovery of synthetic, pre-computed, rotation profiles from their associated rotational splittings.

\subsection{Rotational splittings and rotational kernels}\label{subsec_splitting}
In the case of a slow rotator, the effects of rotation can be treated as a perturbation of the non-rotating spherically symmetric equilibrium state. To the first order the frequencies of the oscillations will be given by
\begin{equation}
\nu_{n,\ell}^m=\nu_{n,\ell}^0+\delta \nu_{n,\ell}^m ,
\end{equation}
where $\nu_{n,\ell}^m$ is the frequency of the modes taking in account rotation and $\nu_{n,\ell}^0$ the frequency of the non-rotating state. The difference between $\nu_{n,\ell}^m$ and $\nu_{n,\ell}^0$ is called the rotational splitting $\delta \nu_{n,l}^m$ \citep{Ledoux1951}. It provides a direct way to determine the internal rotation of stars in asteroseismology. Rotational splittings depend on two quantities, first the rotation profile of the star $\Omega(r)$ and then on the so-called rotational kernel $K_{n,\ell}(r)$. The behaviour of the kernel is directly linked to the physical properties of the oscillation mode with which it is associated. Assuming the rotational profile to be spherically symmetric, a rotational splitting can be expressed mathematically as
\begin{equation}\label{eq_splitting}
\delta\nu_{n,\ell}^m=m\int^R_0  K_{n,\ell}(r) \Omega(r) dr.
\end{equation}
In practice one has such an integral expression for each observed rotational splitting. The information they provide is however degenerate, as it is intimately bound to the nature of the rotation kernel. It is also worth noting they must satisfy mathematical inequalities \citep{Reese2015} and thus each splitting is not mathematically independent of the others. 

Regarding such constraints, Kepler 56 is an excellent target since ten splittings are observed and presented in Appendix~\ref{table_data_sismo}, each splitting carries informations about the rotation profile of the star. However, due to the extreme asymmetry of the splitting at $205.437\ \mu Hz$, we will exclude it from our first order analyses. 

\begin{figure}[h]
\centering
\includegraphics[width=\hsize]{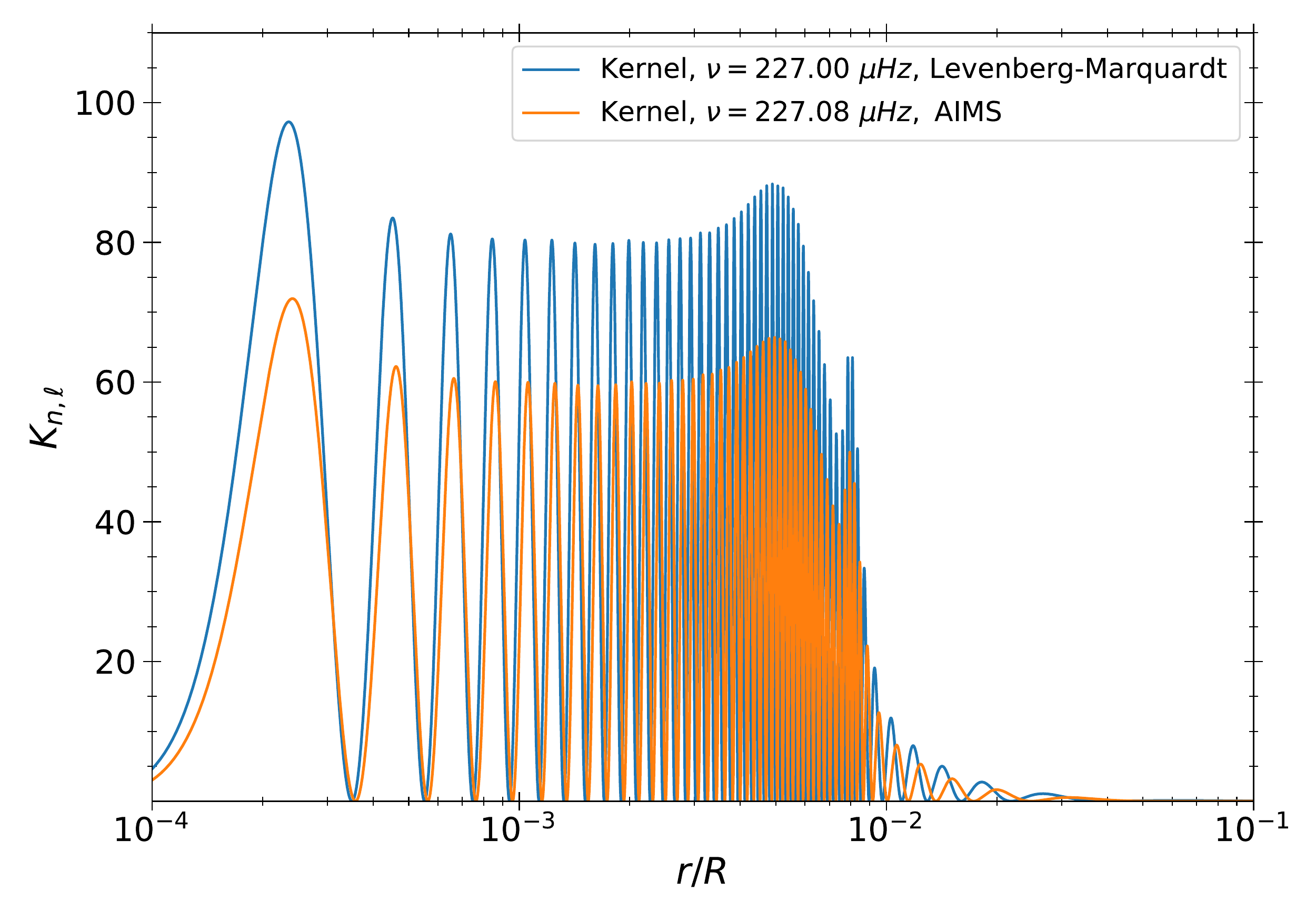}
\caption{Kernels of the same gravity dominated modes from the best fitting model found with AIMS (in orange) fitting the radial modes only and the model found with the Levenberg-Marquardt minimization technique (in blue) taking into account constraints on dipolar mixed modes..}
\label{fig._kernels}
\end{figure}

An important aspect of the modelling of RGB stars is the strong non-linear behaviour of the modes, as a slight shift in frequency may induce a large change in the eigenfunctions and thus a complete misunderstanding of the information carried by the rotation kernels. The non-linearity is controlled in our modelling with direct reproduction of the cavities coupling through the values $\nu_{cross}$ and the verification of the Echelle diagram. Ultimately, the non-linearity can be illustrated by looking directly at the amplitude of kernels, as is shown in Fig.~\ref{fig._kernels} where kernels of a given mode are compared for two different models, one built only fitting radial oscillation modes and the other fitting the whole oscillation spectrum. The clear difference of amplitude can be seen directly impacting their behaviour, illustrating perfectly the structural dependency in the splittings and the importance of taking into account constraints on dipolar mixed modes to obtain a reliable fit of the internal structure before going through a thorough rotational analysis of a red giant star.

The splittings of the whole $\ell=1$ spectrum is illustrated in Appendix~\ref{Appendix_non_linear} for the models obtained with AIMS and the Levenberg-Marquardt minimization technique for a given rotation profile. This figure shows the impact of differences in the rotational kernels, induced by slight differences in the stellar models, directly impacting the splittings. Appendix~\ref{Appendix_non_linear} illustrates perfectly the needs to reproduce the whole oscillation spectrum, here by the means of the Levenberg-Marquardt minimization technique to avoid biased inferences due to an improper reproduction of the oscillation cavities. Finally, this figure shows that the use of the AIMS model could induce a strong variation in the derived surface and core rotation, with in such a case, a slower surface rotation and a faster rotation expected in the core.

\subsection{The MCMC method}\label{subsec_MCMC_method}

Markov chain Monte Carlo (MCMC) denotes a class of algorithms sampling a probability distribution. The result from such an algorithm is, ideally, the posterior probability distribution of the variables considered. To perform our analysis, we used the \textit{emcee} package for python 3 \citep{emcee2013} with the Affine Invariant MCMC Ensemble sampler and the parallel tempering approach implemented in the package \textit{ptemcee} \citep{ptemcee2016, emcee2013}.

The MCMC analysis was based on Eq.~\ref{eq_splitting}, using LOSC and \textit{InversionKit} \citep{InversionKit2016}, we extracted the kernels of individual pulsation modes for the observed splittings presented in Table~\ref{table_data_sismo}. 

Parametric rotation profiles are, then, injected in Eq. \ref{eq_splitting} to get an estimate of the splittings. The role of the MCMC analysis is to optimize the free parameters of the rotation profiles to reproduce the observations. To do so each step of the MCMC is decomposed as follows:
\begin{itemize}
\item For each walker, splittings are computed following Eq. \ref{eq_splitting} using the previously extracted kernels and a rotation profile computed with a given set of parameters. 
\item The splittings obtained by integration are then compared to the observed ones by means of the logarithm of the likelihood $\mathcal{L}$ :
\begin{equation}
\ln \mathcal{L} = -\frac{1}{2}\chi^2,
\end{equation}
with
\begin{equation}\label{eq_chi2}
\chi^2=\sum_i \left(\frac{\delta\nu_{model}-\delta\nu_{obs}}{\sigma_{\delta\nu_{obs}}}\right)^2,
\end{equation}
where $\sigma_{\delta\nu_{obs}}$ are the uncertainties on the observed splittings.
\item New parameters of the rotation profile are then adopted, based on the new value of $\ln \mathcal{L}$, and the loop starts back from its first step until the number of iterations requested is achieved. 
\end{itemize}
The results obtained at the end correspond to the median of the parameter distributions, the errors correspond to the first $15.9\%$ and $84.1\%$ of the distributions. The viability of the MCMC approach is verified through the computation of the autocorrelation for each free parameters and an analysis of the walkers trajectories. The walkers correspond to chains of stochastic processes randomly sampling the probability distribution according to the algorithm.
\\Since we are probing a highly degenerate parameter space we implemented a parallel tempering approach with \textit{ptemcee} and \textit{emcee} to treat these complex problems. We based the number of iterations N on the observed autocorrelation of each sample to guarantee that the autocorrelation of all the MCMC results presented below are at least around $N/50$  as recommended by \cite{emcee2013} to have acceptable sampling. The number of walkers, temperatures and the ladder of temperature were chosen in order to reduce the autocorrelation, increase the overall quality of the probability distribution obtained, reducing the required number of iterations and the MCMC running time. In the following section we detail the technical characteristics of each MCMC run. 

\subsection{Confirmation/Verification of the method }\label{subsec_MCMC_confirmation}
Before applying this method to the actual target, Kepler 56, its robustness should be investigated. To do so, we carried out tests on artificial data using the following methodology
\begin{itemize}
\item First, the viability of the integrator is checked by comparing the splittings found by our integrator to the one computed with \textit{InversionKit} for a given rotation profile.
\item Synthetic splittings are, then, created with a rotation profile for a given set of parameters. The MCMC is run as described in Sect. \ref{subsec_MCMC_method}. The objective of this setup is to prove that the MCMC is capable to recover the parameters of the rotation profile used to build the synthetic splittings. 
\end{itemize}
The results obtained on such tests depend on the parametric rotation profile used and especially the number of input free parameters. Assuming a large number of free parameters may well lead to degeneracies as the information on the rotation profile given by the splittings is limited. Thus, the degree of "customization'' of the synthetic rotation profiles is intrinsically limited by the number of observed splittings and by the nature of the modes for which they are observed. In our analysis, we tailor the tests to the case of the dataset of Kepler 56, meaning that we use the exact same modes, with their observational uncertainties to prove that the method is viable for this particular target.

The parametric profiles used find some basis in more physical analyses of AM transport processes, while trying to keep the number of free parameters as limited as possible. From example, we introduce the simple following definition
\begin{equation}\label{eq_profile_power_law}
\Omega(r)=
	\left\lbrace
		\begin{aligned}
		\ &\Omega_{\rm{core}} \qquad & r \leq r_{BCE},
		\\ \ &\Omega_{\rm{core}}\left(\frac{r_{BCE}}{r}\right)^\alpha \qquad & r \geq r_{BCE},
		\end{aligned}
	\right.
\end{equation}
where $r_{BCE}$ is the radius at the base of the convective envelope, $\alpha$ and $\Omega_{\rm{core}}$ are the two free parameters of this profile. In our following analysis, we will refer to this rotation profile as the "power law" profile.
\\In this case the rotation is assumed as solid-body in the radiative zone and then differential in the convective envelope, following a functional dependency in $r^{-\alpha}$. The particular case of $\alpha=1$ corresponds to the prescription of \cite{Kissin2015} and \cite{Takahashi2020} regarding the expected rotation profile in stars with large-scale fossil magnetic fields. 
\\For the MCMC analysis, this profile shows no degeneracies thus the Affine Invariant MCMC Ensemble sampler of \textit{emcee} is extremely efficient to sample the parameter space. Each MCMC analysis with the power law rotation profile uses 20 walkers for 5000 iterations and a burn-in of 200 iterations. The priors are uniform for both $\alpha$ and $\Omega_{\rm{core}}$. The free parameter $\Omega_{\rm{core}}$ is assumed to be positive. Depending on the cases studied $\alpha$ is limited differently, during MCMC check-up $\alpha$ is explored form $0$ to $10$ and for the study of Kepler-56 the limits of $\alpha$ are detailed in Section~\ref{subsec_MCMC_results}.

\begin{figure}[h]
\centering
\includegraphics[width=\hsize]{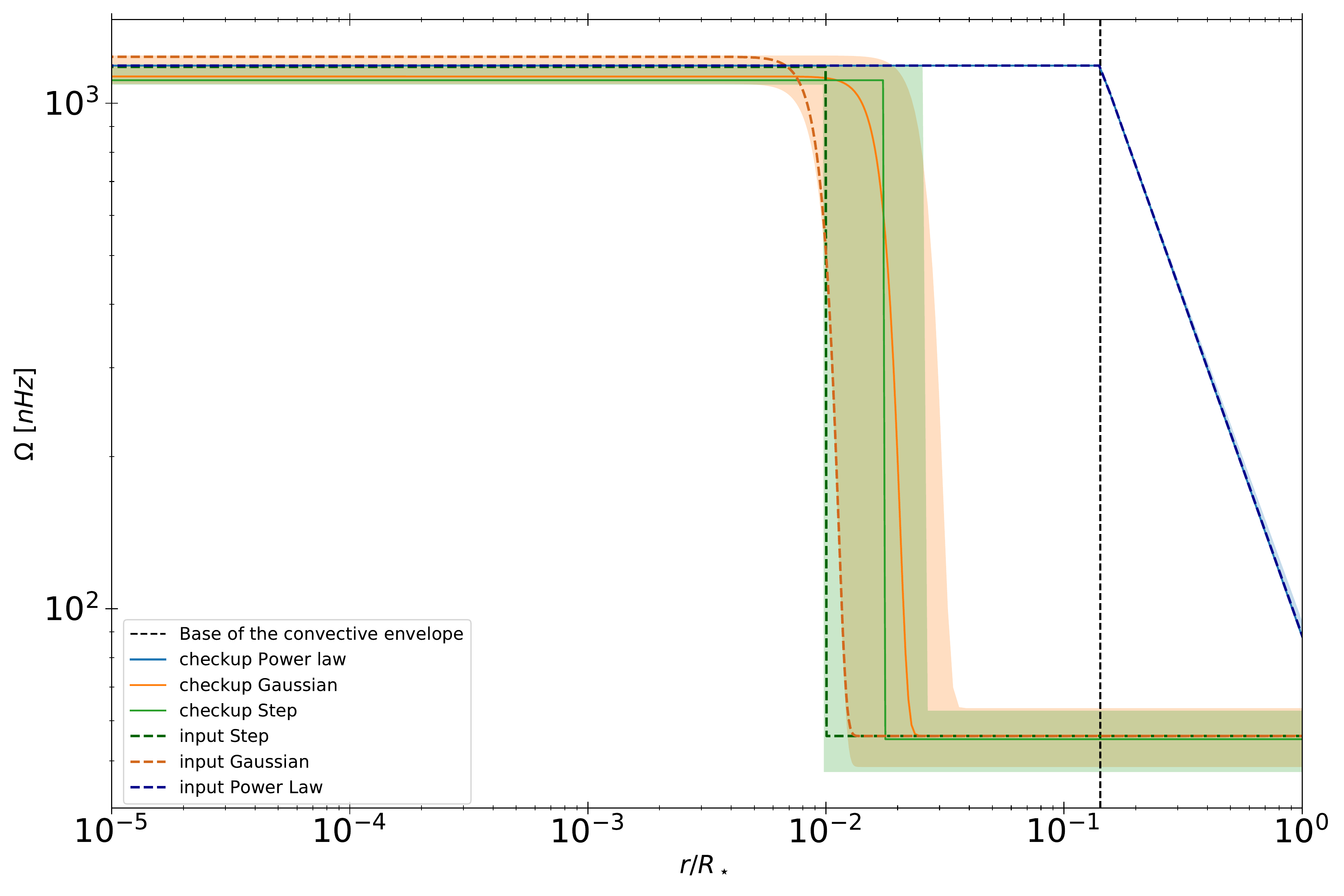}
\caption{Illustration of the different rotation profiles found with the MCMC analysis (in color), their uncertainties as a shaded area and the rotation profiles they should replicate in black. }
\label{fig._checkup}
\end{figure}

A prescription aimed at mimicking the profiles of \cite{Eggenberger2012} is also part of our investigation through the following rotation profile
\begin{equation}\label{eq_profile_gaussian}
\Omega(r)=(\Omega_{\rm{core}} - \Omega_{\rm{surf}}) \exp\left(-\left(\frac{1}{\sigma} \frac{r}{r_{norm}}\right)^8\right) +\Omega_{\rm{surf}},
\end{equation}
with $r_{norm}$ an arbitrary normalisation constant $r_{norm}=0.00804$ , $\Omega_{\rm{core}}$, $\Omega_{\rm{surf}}$ and $\sigma$ the free parameters. In our analysis we will refer to this rotation profile as the "gaussian" profile. 
\\This profile predicts a differential rotation in the radiative zone with a solid-body rotation in the envelope due to the highly efficient transport of AM by convection. 
\\For the MCMC analysis of the gaussian profile, we used uniform priors on each free parameter. Due to the highly degenerate parameters space and the highly multimodal expected posterior distribution we used the parallel tempering approach of \textit{ptemcee} with 40 walkers, 8 temperatures, 2000 iterations and 100 steps of burn-in for each results of the gaussian rotation profile presented below. All the positive values of the surface rotation $\Omega_{\rm{surf}}$   and  $(\Omega_{\rm{core}} - \Omega_{\rm{surf}})$ were explored to study all the possible scenarios, and in particular discard solid-body rotation. Finally, with $\sigma$ we explored a transition in the central area of the star by setting its minimum to $0.01$, to avoid divergences, and its maximum to $4$ expecting a transition in the rotation profile close the chemical composition gradient following \cite{Eggenberger2012}. We also do not expect to have any constraints after $\sigma=4$ because the splittings of $\ell=1$ modes are not affected by the rotation of this region, only observations of higher degrees splittings can give hints of the rotation for the rest of the radiative zone. 
\\The position of the transition from the rapid internal to the slow surface rotation is of particular interest for this profile. We expect an abrupt transition, due to the important chemical composition gradient located at the peak of the Brunt-Vaïsälä frequency close to the hydrogen-burning shell. Such a behaviour is expected from a variety of physical mechanisms, bound to a turbulent nature, such as magnetic instabilities \citep{Eggenberger2019} and internal gravity waves \citep{Pincon2017} that would be inhibited by the effects of mean molecular weights gradients. 
\\In this context we tried to characterise the slope of the transition with several rotation profiles. Finally, we choose to present a last simple rotation parametrisation as
\begin{equation}\label{eq_profile_step}
\Omega(r)=
	\left\lbrace
		\begin{aligned}
		\ &\Omega_{\rm{core}} \qquad & r \leq r_{\rm{TR}},
		\\ \ &\Omega_{\rm{surf}} \qquad & r >r_ {\rm{TR}},
		\end{aligned}
	\right.
\end{equation}
where $\Omega_{\rm{core}}$, $\Omega_{\rm{surf}}$ and $r_{TR}$ are the free parameters. In our analysis, we will refer to this rotation profile as the "step" profile.
\\This profile models the transition from core to surface rotation as a discontinuity to ensure the expected abrupt transition. 
\\ For the MCMC analysis of the step rotation profile, uniform priors are used for all free parameters. As in the gaussian rotation profile, the parameter space shows a strong degeneracy between the parameters thus we use the parallel tempering implemented by \textit{ptemcee} to properly sample the prior distribution. For each MCMC run with the step rotation profile we used $40$ walkers, 8 different temperatures and 2000 iterations with a burn-in of 100 steps. All the positive values $\Omega_{\rm{surf}}$ and $\Omega_{\rm{core}}$ were explored. Finally, $r_{\rm{TR}}$ is explored from $0.001$ to $0.03 R_\star$ expecting a transition in the rotation profile close the chemical composition gradient following \cite{Eggenberger2012}. 
\\The objective with the last two rotation profiles is to define whether the slope of the transition should be close to a discontinuity or much smoother.
Other profiles were tested, but either included too many free parameters and led to degeneracies in the solutions, or poor convergence. As a result, we decided to limit ourselves to the ones given by Eqs. \ref{eq_profile_power_law}, \ref{eq_profile_gaussian} and \ref{eq_profile_step}.For the chosen rotation profiles the solution from the MCMC check-up as well as the rotation profile they should recover is shown in Fig.~\ref{fig._checkup}.

The MCMC check-up shows no degeneracy for the parameters of the power law function and recovers easily such a profile. The step and gaussian function are able to recover the input rotation profile within their error bars even if a degeneracy appears between the parameters controlling the transition and the core rotation. For all the rotation profiles tested, the surface rotation in particular is always found without any degeneracy, thus, this quantity is very well constrained by our method.

\begin{figure}[h] 
\centering
\includegraphics[width=\hsize]{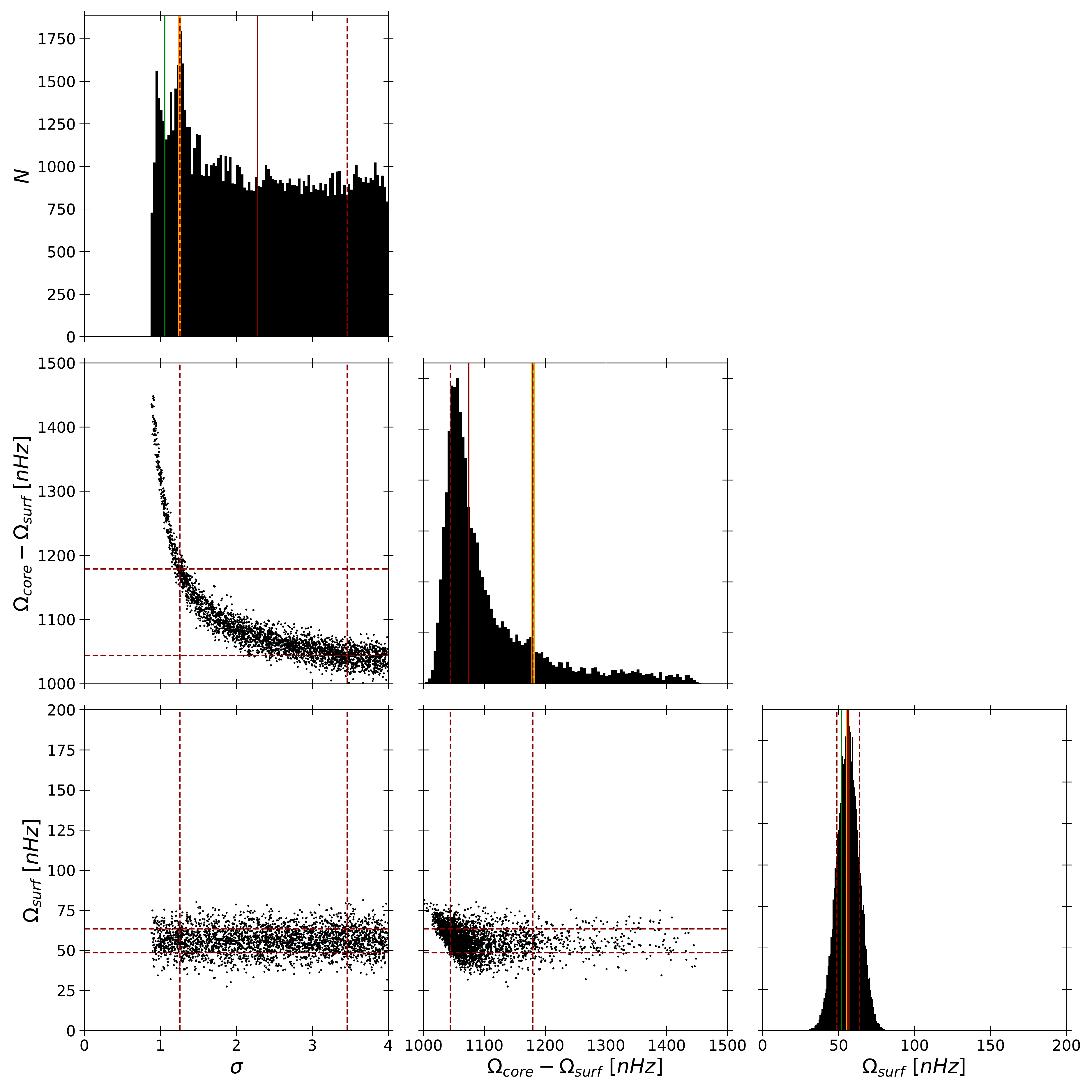}
\caption{Parameter probability distribution obtained with the check-up of MCMC analysis on the gaussian function. The red solid line represents the median of the distribution, the red dotted lines correspond to the $1-\sigma$ uncertainties assuming a Gaussian distribution. The solid green line represent the maximum of the likelihood solution while the solid orange correspond to the initial input parameters that the check-up should recover. Only a small fraction of the sampler is displayed to avoid overweighted figures, however, the histograms were constructed with the full sample.} 
\label{fig._checkup_gaussian}
\end{figure}

With the step and gaussian rotation profile the limit of the method presented here can already be seen for a star without observed $\ell=2$ splittings like Kepler 56. One could expect for both rotation profiles a relationship or a correlation between the position of the transition and the core rotation. However, as illustrated in the middle left panel of both Fig.~\ref{fig._checkup_gaussian} and Fig.~\ref{fig._checkup_gate} representing the posterior distribution in check-up for respectively the step and gaussian rotation profile, after $\sigma=1.5$ or $r_{\rm{TR}}= 0.012 R_\star$ the position of the transition and the core rotation are close to be independent parameters. This independence is probably caused by a lack of sensitivity in the kernels after $\sigma=1.5$ or $r_{\rm{TR}}= 0.012 R_\star$, the amplitude of the kernels is not sufficient to really impact the splittings thus this area of the parameter space is not constrained and the MCMC can sample freely this region finding the same solution, independently from the position of the transition. The solution obtained after $\sigma=1.5$ or $r_{\rm{TR}}= 0.012 R_\star$ is at an almost constant $\Omega_{\rm{core}}$ meaning that a major peak in the distribution is obtained around this value, as seen in the middle panel of Figures~\ref{fig._checkup_gaussian} and~\ref{fig._checkup_gate}. This peak is a  feature of the uniform sampling made in the region with the lack of resolution and thus it has no real physical justification. 

For the gaussian rotation profile a second peak with a lower amplitude located around the maximum likelihood and initial input parameters can be seen in the middle panel of Fig.~\ref{fig._checkup_gaussian}, this peak and its equivalent in the distribution of $\sigma$ shows that the input parameters can be recovered at the limit of the one sigma error bars if the exploration of the parameter space is dominated by a region where the kernels have no resolving power.
\\The posterior distribution of the step rotation profile is very similar to the posterior distribution of gaussian rotation profile. The input parameters are also recovered within the errors bars even if the distributions shows signs of multimodality especially for $\Omega_{\rm{core}}$. The multimodality observed in the posterior distribution can be attributed to the behaviour of the rotation kernels and the shape of the rotation profile. Namely, the sharp, discontinuous transition in the rotation profile will be placed at the location where it has a maximum impact on the splittings, near the local maximum of the kernels. 

As mentioned before, the surface rotation is well constrained by our method, the posterior distribution of the surface rotation, illustrated in the bottom right panel of Fig.~\ref{fig._checkup_gaussian} and Fig.~\ref{fig._checkup_gate}, shows a single gaussian type peak centered in the input parameters. 

\section{Kepler 56}\label{Kepler56}
The MCMC analysis was run with the splittings presented in table \ref{table_data_sismo}. We are now going to detail the results obtained for each rotation profiles in Sect. \ref{subsec_MCMC_results}. We then discuss the implication of this analysis in the context of a  missing AM transport process during the evolution of the star in Sect. \ref{subsec_AM}.
\subsection{Results from the MCMC analysis}\label{subsec_MCMC_results}

As a starting point to the analysis, we carried out a standard inversion procedure using the Substractive Optimally Localized Averages method \citep[SOLA,][]{Pijpers1994} to derive the core rotation of Kepler 56, to have a measurement "independent" from the MCMC analysis. We derived an internal rotation in the core of Kepler 56 of $1100 \pm100$ n$\rm{Hz}$ from the SOLA inversion. This results will be compared to the final result of the MCMC analysis, keeping in mind that both techniques take very different approaches and provide complementary views on the inversion problem.

Before presenting the actual results of the MCMC analysis some additional information should be given on the rotation profiles and the final probability distributions of their parameters.

The power law profiles were separated in two cases, a first MCMC analysis, called "Limited power law", was done limiting the free parameter $\alpha$ to $1.5$ following the prescription of \cite{Kissin2015, Takahashi2020} expecting that $\alpha$ lies between $1$ and $1.5$. The second MCMC analysis on this profile, called "Unlimited power law", lets all the free parameters unlimited in their values. In both cases the final distribution of parameters show no degeneracy and a well defined solution, the parameter probability distribution of the unlimited power law function is illustrated in Fig.~\ref{fig._checkup_gate}. Both best fitting MCMC results and median of the distribution are presented in Table~\ref{table_MCMC_results} for both rotation profiles.

\begin{figure}[h]
\centering
\includegraphics[width=\hsize]{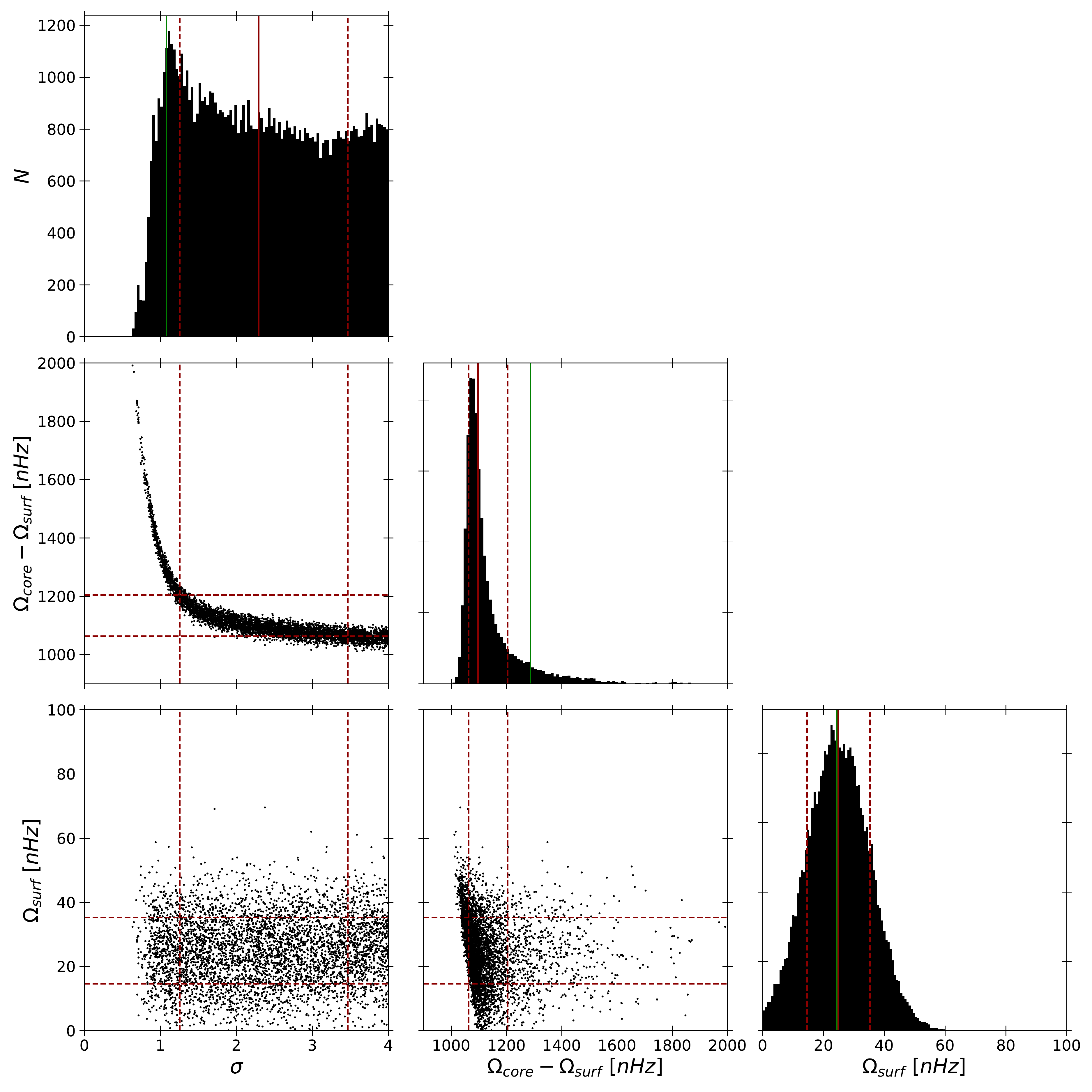}
\caption{Triangle diagram presenting the posterior distribution from the MCMC analysis on the gaussian function with the data of Kepler 56. The red solid line represents the median, the red dotted lines it's one-sigma uncertainties and the green solid line the maximum of the likelihood solution. Only a small fraction of the sampler is displayed to avoid overweighted figures, however, the histograms were constructed with the full sample.} 
\label{fig._MCMC_distri_gaussian}
\end{figure}

As mentioned in Sect.~\ref{subsec_MCMC_confirmation}, the gaussian rotation profile shows a correlation between the parameters controlling the position of the transition and the core rotation. The posterior probability distribution obtained using Kepler 56 data, illustrated in Fig.~\ref{fig._MCMC_distri_gaussian}, shows the exact same general behaviour as the checkup. A clear peak in the probability of the parameter controlling the core rotation can be seen while the position of the transition is poorly constrained.
The final parameters values obtained from the MCMC analysis are presented in Table~\ref{table_MCMC_results}.

\begin{figure}[h]
\centering
\includegraphics[width=\hsize]{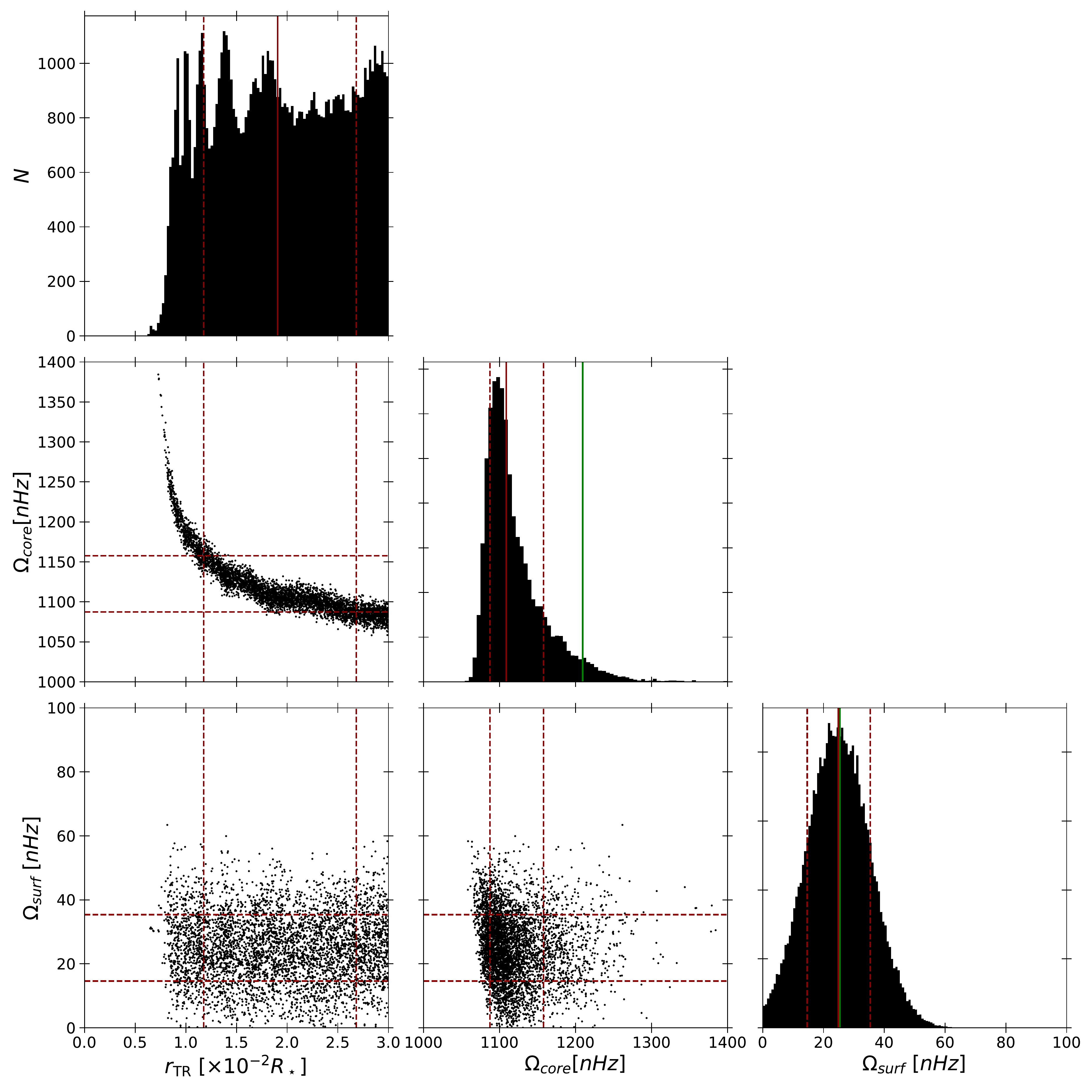}
\caption{Triangle diagram presenting the result from the MCMC analysis on the step function with the data of Kepler 56. The probability distribution of the position of the transition is cut at $0.030\ R$ due to negligible amplitude of kernels after this mark. The red solid line represents the median, the red dotted lines it's one-sigma uncertainties and the green solid line the maximum of the likelihood solution. Only a small fraction of the sampler is displayed to avoid overweighted figures, however, the histograms were constructed with the full sample.} 
\label{fig._MCMC_distri_gate}
\end{figure}

The last rotation profile studied is the step rotation profile. The posterior parameters probability distributions obtained with Kepler 56 data is illustrated in Fig.~\ref{fig._MCMC_distri_gate}. The distributions show a strong degeneracy with several couples of parameters maximizing the likelihood. As expected, the surface rotation is efficiently constrained independently from the core rotation or the position of the transition.

\begin{table*}[h]
\centering
\caption{Results of the MCMC analysis for each rotation profiles. The values between parenthesis indicate the parameters deduced from the free parameters. The $\chi^2$ is computed with Eq. \ref{eq_chi2}. }\label{table_MCMC_results}
\begin{tabular}{llllll}
\hline\hline
\multirow{2}{*}{Model parameters}       & \multirow{2}{*}{Limited Power Law} & \multirow{2}{*}{Unlimited Power Law} & \multirow{2}{*}{Step}  & \multirow{2}{*}{Gaussian} \\
                                        &                                    &                                      &                       &                           \\ \hline
\multirow{2}{*}{$\Omega_{\rm{core}}\ [$n$\rm{Hz}]$} & \multirow{2}{*}{$1005^{+4}_{-4}$}                 & \multirow{2}{*}{$1057^{+8}_{-8}$}                   & \multirow{2}{*}{$1081^{+16}_{-27}$}    & \multirow{2}{*}{$1096^{+107}_{-34}$}        \\
                                        &                                    &                                      &                       &                           \\
\\$\Omega_{\rm{surf}}\ [$n$\rm{Hz}]$               & ( $54.4^{+0.1}_{-0.1}$ )                                 & ( $0.146^{+1.502}_{-0.145}$ )                                     & $25.9^{+14.7}_{-13.3}$ 	                   & $24.8^{+10.5}_{-10.2}$                         \\
\\$\alpha$                                & $1.494^{+0.004}_{-0.009} $                                  & $4.55^{+2.86}_{-1.24} $                                  & \textunderscore                     & \textunderscore                     \\
\\$\sigma$                                & \textunderscore   & \textunderscore                                 & \textunderscore                                                        & $2.29^{+1.17}_{-1.04}$                         \\
\\$r_{TR}$                               & \textunderscore                                  & \textunderscore                                    & $0.0191^{+0.0076}_{-0.0073}$                      & \textunderscore                         \\ \\ \hline
\multirow{2}{*}{$\chi^2$}               & \multirow{2}{*}{$90.2$}                 & \multirow{2}{*}{$5.98$}                   & \multirow{2}{*}{$2.64$}    & \multirow{2}{*}{$2.72$}        \\
\\ \multirow{2}{*}{$\rm{BIC}$}               & \multirow{2}{*}{$51.7$}                 & \multirow{2}{*}{$9.58$}                   & \multirow{2}{*}{$8.46$}    & \multirow{2}{*}{$8.92$}        \\
                                        &                                    &                                      &                       &                           \\ \hline
 
\end{tabular}
\end{table*}

For the sake of completeness, we also compare the values of the Bayesian Information Criterion (BIC) indicator, defined as
\begin{equation}
\rm{BIC}= k\ln(n)-2\ln(\widehat{\mathcal{L}})
\end{equation}
where $k$ is the number of free parameters, $n$ the number of constraints, $\widehat{\mathcal{L}}$ the maximum of the likelihood.

\begin{figure*}[h]
\centering
\includegraphics[width = 13cm]{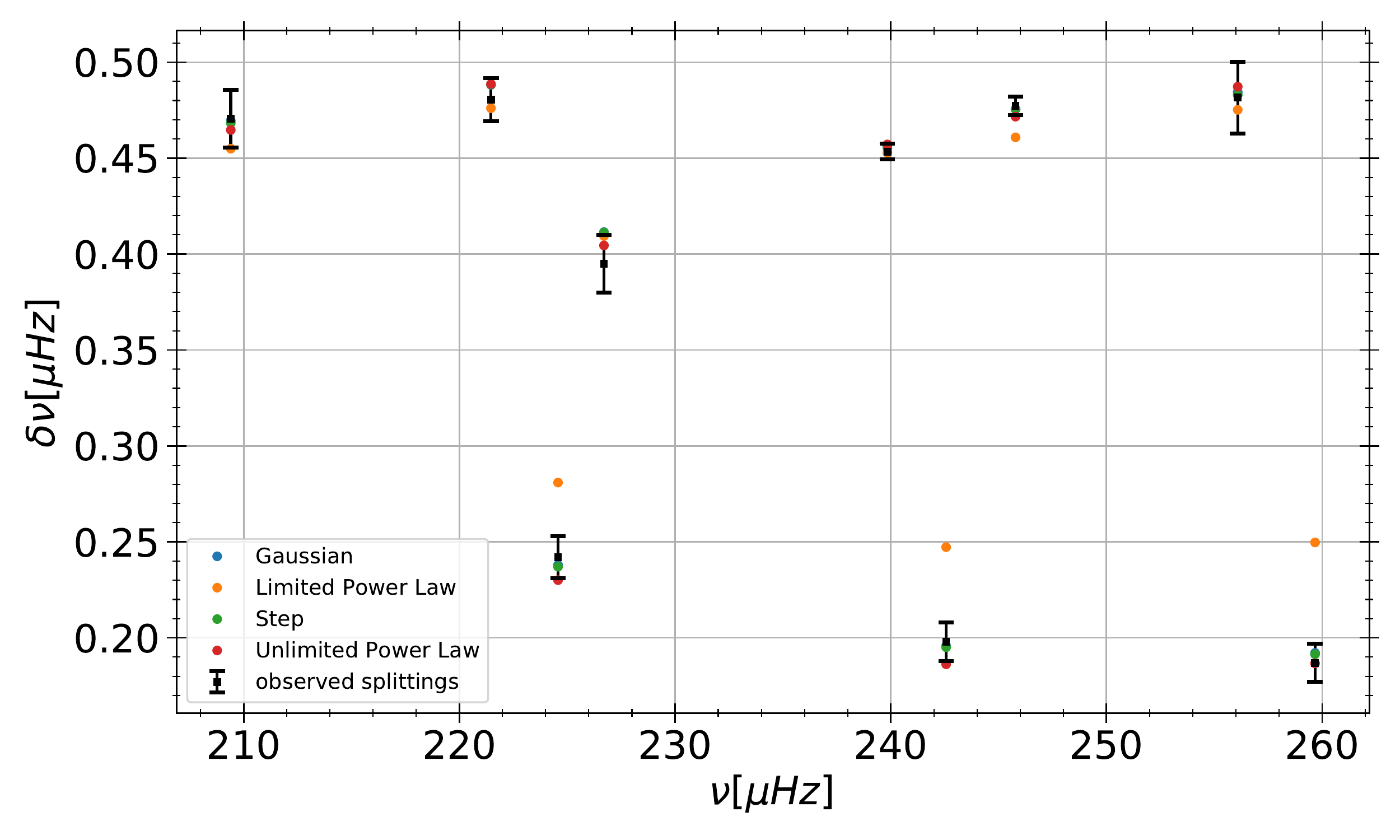}
\caption{Splittings observed and presented in Tab. \ref{table_data_sismo} in black and compared to the modelled splittings with the different rotation profiles obtained with the median of the MCMC final distribution of parameters.}
\label{fig._splittings}
\end{figure*}

The distribution obtained for the limited power law shows that no clear minimum is found in the parameter space allowed. 
This rules out this profile as a potential solution for the case of Kepler 56, as we showed with synthetic data that it should be possible to retrieve such a profile from the available dataset of Kepler 56. For the unlimited case a minimum is found with a $\chi^2=5.98$. In this case, the minimisation algorithm tends to compensate the advanced position of the transition by making a sharper rotation profile in the convective envelope. The resulting rotation profile is physically unlikely due to the extremely slow rotation found at the surface and the strong dependency in $\sim r^{-4.55}$ of the rotation in the convective zone, in disagreement with a local conservation of AM. All modelled splittings for each of the parametric profiles studied are presented and compared to the observed ones in the Fig. \ref{fig._splittings}.

On the other hand, the step and gaussian rotation profiles reproduce with more accuracy the observed splittings with respectively $\chi^2=2.64$, $\chi^2=2.72$ and a $\Delta \rm{BIC} \approx 40$ compared to the power law rotation profile. Despite the clear $\chi^2$ and $\rm{BIC}$ advantage of these profiles, the introduction of a third free parameter controlling the position of the transition also leads to a degeneracy between the free parameters as illustrated in Figures~\ref{fig._MCMC_distri_gaussian} and \ref{fig._MCMC_distri_gate}. Even with this degeneracy, major peaks in the parameter probability distribution came out of the noise. 

Looking at the posterior probabilities, we observe a similar behaviour as in our checkups using artificial data. This leads to believe that a strong degeneracy exists between the various parameters of the profiles we tested using our MCMC approach. This resolution limit is a direct result of the behaviour of the rotation kernels, that show low amplitudes above $0.03R_\star$, meaning that the rotation profile of Kepler 56 cannot be constrained from the rotation splittings above this limit without additional independent constraints. 

In the specific case of Kepler 56, the surface rotation determined from starspots could have played such a role. However, we find it to be very precisely constrained by the MCMC technique, at a mean value of $25.4^{+9.0}_{-8.4}$ nHz. We should mention that the $\Omega_{\rm{surf}}$ found with our MCMC analysis corresponds to the rotation in the acoustic cavity. The presence of a shear layer at the surface of the star or a breaking of the spherical symmetry in the rotation profile in the envelope can impact what we deem as ``surface rotation'' and has to be considered when comparing to measurements from other techniques.
\cite{Huber2013} reported a stellar surface rotation with the value of $156\pm6$ nHz. Using $156$ nHz as input for our analysis, we were unable to reproduce the ratio of the p and g dominated dipolar modes, as already noted by \cite{Klion2017}. However, while they could reproduce the minimum ratio of the splittings with a surface rotation of $78$ nHz, we find that we require an even slower rotation. The origin of these differences could be found in the fact that we carried out a detailed structural modelling of Kepler 56 and analyzed the entire oscillation spectrum, while they restricted themselves to the minimum ratio of the g and p dominated splittings. 
\cite{Huber2013} discussed the value they obtained, stating that the period is very close to the duration of the observation quarters of the \textit{Kepler} spacecraft and that the value given could be an harmonic of the actual rotational frequency. As noted above, our measurement cannot be directly compared to that of \cite{Huber2013}, as other effects such as a breaking of spherical symmetry of rotation in the acoustic cavity or a surface shear layer could reduce the disagreements with the value obtained from starspots by \cite{Huber2013}.
The strong discrepancy between core and surface rotation is seen in the ratios of the splittings of p-dominated and g-dominated modes, as discussed in \citep{Eggenberger2012} and visible in our case in Fig.\ref{fig._splittings} can also support this result.

Meanwhile, our MCMC technique provides a core rotation value that is in agreement with the SOLA inversion results as well as the value provided by \cite{Klion2017}. Indeed, the large number of g dominated modes strongly constraints this parameter to a narrow range of values in our modelling. Had the surface rotation not been so widely different, we could have hoped to more precisely locate the position of the transition in rotation in the radiative zone.


The observed splittings as well as the ones modelled with the gaussian and step rotation profiles are illustrated in Fig.~\ref{fig._splittings}. 

\begin{figure}[h]
\centering
\includegraphics[width=\hsize]{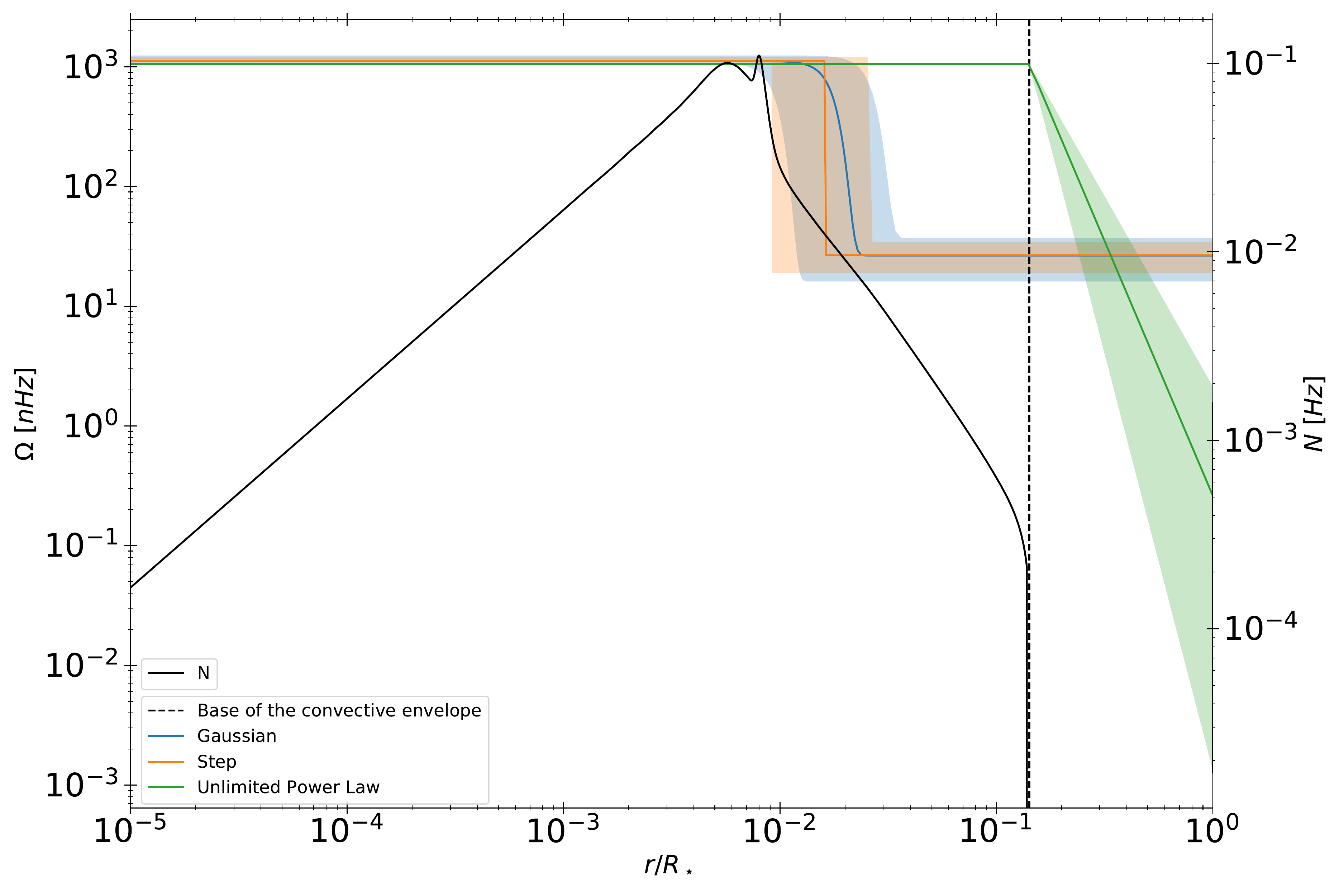}
\caption{Three different profiles obtained with the median of the MCMC final distribution of parameters compared to the Brünt-Väisälä frequency in black. The dotted lines represent the best MCMC solutions for each rotation profile with the Kepler 56 data.}
\label{fig._all_rotation_profiles}
\end{figure}

We note that all the observed splittings illustrated in Fig.~\ref{fig._splittings} are reproduced within their one-sigma uncertainty by both the step and the gaussian rotation profile. The robustness of our approach with respect to issues in the observational dataset was also tested. We modified the splitting at $\nu=221.464$ nHz of about $5\%$ and checked whether the results obtained with our method were impacted by this modification. We found our method to be less sensitive to such issues, unlike classical inversion techniques such as the SOLA method \citep{Pijpers1994, Pijpers1997}.
The final rotation profiles associated with the splittings presented in Fig.~\ref{fig._splittings} are illustrated in Fig.~\ref{fig._all_rotation_profiles}.

\subsection{Impact on the missing AM transport process}\label{subsec_AM}

Characterizing the rotation profile of evolved stars is of course of prime importance to determine the physical nature of the efficient AM transport process needed to correctly reproduce the asteroseismic measurements available for these stars. The present results on the internal rotation of Kepler 56 show that a rigid rotation in the radiative interior followed by a power law transition in the convective envelope, as theorized by \cite{Kissin2015,Takahashi2020} is incompatible with the seismic observations. Our findings seem to favour a transition possibly located in the deep radiative layers. While our results discard large-scale magnetic fields imposing a rotation as prescribed in \cite{Kissin2015,Takahashi2020}, they tend to favour AM transport by magnetic instabilities, that are strongly affected by the presence of steep chemical composition gradients \citep[e.g.][]{denHartogh2019,Eggenberger2019II}, here located at the hydrogen-burning shell. Indeed, our parametric profiles mimicking the results of evolutionary computations using such processes reproduce very well the observed rotational splittings of Kepler 56. However, the quality of the data does not allow to very precisely locate the transition in the radiative zone in the rotation profile, as a result of the uncertainties on the surface rotation measured from starspots.

\section{Conclusion}\label{Conclusion}

In this study, we carried out a detailed analysis of the properties of Kepler 56, a well-known early red giant branch exoplanet-host star observed by \textit{Kepler} \citep{Huber2013}. Our goal was to study the internal rotation of the star using seismic data, and see if we could provide constraints on the physical nature of the missing AM transport mechanism acting inside post main-sequence stars. To do so, we needed to discriminate between rotational profiles showing a solid-body rotation in the whole radiative region from profiles showing a transition in rotation in these layers.

We started by carrying out an extensive seismic modelling procedure of the structure of Kepler 56. We combined global minimization techniques with the AIMS software \citep{AIMS2016, AIMS2019}, local minimization following the approach of \citet{Deheuvels2011}, and seismic inversions following \citet{Reese2012den, Buldgen2019}. This led us to obtain an excellent agreement with the seismic and non-seismic constraints, and allowed us to consider the modelling robust enough to carry out an extensive analysis of the rotational properties of the star. We started by testing our methodology on synthetic data with the exact same number and type of oscillation modes as for Kepler 56, as well as the same uncertainties on the rotational splittings. The results of this analysis proved the robustness of the method in the considered case and gave us confidence in applying our technique to the real target. We then show that we are able to discriminate between various types of rotation profiles and reject the hypothesis of solid-body rotation in the whole radiative zone of Kepler 56 followed by a slowly decreasing power law profile in radius, as theorized by \cite{Kissin2015,Takahashi2020}. Our analysis shows that a transition in the rotation profile located in the radiative region, close to the hydrogen burning shell is favoured. Indeed, parametric models including such a sharp transition close to the shell are able to reproduce the observed splittings of Kepler 56 with an excellent agreement. The stellar envelope rotation is efficiently constrained to a value of $25.4^{+9.0}_{-8.4}$ n$\rm{Hz}$, which is a factor $6$ lower than the surface rotation suggested by starspot measurements reported by \cite{Huber2013}. A shear layer at the surface of the star, or a breaking of the spherical symmetry of the rotation profile in the convective envelope, as in the Sun, could potentially reduce this discrepancy.

By ruling out the expected profiles in the case where large-scale magnetic fields would be ensuring the angular momentum transport, we do not support their presence as potential solutions for the current observed discrepancies between theoretical and observed rotation properties of red giants. This however does not mean that they could not play an important role in peculiar cases, but there appears to be at least another process at play. Such a process could be found in magnetic instabilities, that would lead to sharp transitions in the internal rotation profile located in regions of steep chemical composition gradients inside the star.


The main limitation of the present method is that it requires very high quality data, a large number of observed splittings, as well as an extensive modelling of the stellar structure. Moreover, our method is limited by the parametric description of the rotation profiles inside the star. However, it is definitely applicable to some of the best \textit{Kepler} targets \citep[see e.g.][]{DiMauro2016, DiMauro2018}, and also potentially to TESS targets with long observation durations, some of which having even better datasets than Kepler 56.

\begin{acknowledgements}
G.B. acknowledges fundings from the 
SNF AMBIZIONE grant No. 185805 
(Seismic inversions and modelling of transport processes in stars).

 P.E. and S.J.A.J.S. have received 
funding from the European Research Council (ERC) 

under the European Union's Horizon 2020 research and innovation programme 
(grant agreement No 833925, project STAREX).
A.M. and J.M. acknowledge support from the European Research Council 
(ERC grant  agreement No. 772293 for the project ASTEROCHRONOMETRY). 
This article used an adapted version of InversionKit, a software developed 
within the HELAS and SPACEINN networks, funded by the European Commissions's 
Sixth and Seventh Framework Programmes.

\end{acknowledgements}

\newpage
\bibliography{KEP-56_biblio.bib}

\newpage
\appendix
\section{Seismic data}

\begin{table}[h]
\caption{Seismic data obtained by \cite{Huber2013} and used in this article. }
\begin{tabular}{ccc}\label{table_data_sismo}
Degree $\ell$        & Frequency $\nu\ [\mu \rm{Hz}]$ & Splitting $\delta\nu\ [\mu \rm{Hz}]$ \\ \hline
2                    & $196.888 \pm 0.019$       & \textunderscore                               \\
0                    & $198.985 \pm 0.017$       & \textunderscore                               \\
2                    & $214.017 \pm 0.025$       & \textunderscore                               \\
0                    & $216.237 \pm 0.016$       & \textunderscore                              \\
2                    & $231.654 \pm 0.019$       & \textunderscore                               \\
0                    & $233.760 \pm 0.015$       & \textunderscore                               \\
2                    & $249.135 \pm 0.017$       & \textunderscore                               \\
0                    & $251.150 \pm 0.016$       & \textunderscore                               \\
2                    & $266.617 \pm 0.019$       & \textunderscore                               \\
0                    & $268.683 \pm 0.020$       & \textunderscore                               \\
\multicolumn{1}{l}{} & \multicolumn{1}{l}{}      & \multicolumn{1}{l}{}            \\
1                    & $190.525 \pm 0.023$       & \textunderscore                               \\
1                    & $192.402 \pm 0.029$       & \textunderscore                               \\
1                    & $205.437 \pm 0.009$       & $0.378\pm0.0106$                 \\
1                    & $207.730 \pm 0.025$       & \textunderscore                               \\
1                    & $209.463 \pm 0.011$       & $0.4705 \pm 0.018$              \\
1                    & $221.464 \pm 0.012$       & $0.4825 \pm 0.0111$             \\
1                    & $224.5730 \pm 0.0077$     & $0.242 \pm 0.011$               \\
1                    & $226.7030 \pm 0.0067$     & $0.395 \pm 0.015$               \\
1                    & $239.8421 \pm 0.0034$     & $0.4534 \pm 0.0040$             \\
1                    & $242.5669 \pm 0.0102$     & $0.198 \pm 0.010$               \\
1                    & $245.7844 \pm 0.0039$     & $0.4773 \pm 0.0048$             \\
1                    & $256.086 \pm 0.017$       & $0.4815 \pm 0.0261$             \\
1                    & $259.6717 \pm 0.0086$     & $0.187 \pm 0.010$               \\
1                    & $277.538 \pm 0.024$       & \textunderscore                               \\
\multicolumn{3}{l}{}                                                               \\
\multicolumn{3}{l}{Large separation $\Delta\nu=17.431\pm0.0011\ \mu Hz$}              \\
\multicolumn{3}{l}{Frequency of max. power $\nu_{max}=244.3\pm 1.3\ \mu Hz$}      
\end{tabular}
\end{table}

\section{Illustration of the non-linear behaviour in the stellar modelling}\label{Appendix_non_linear}

\begin{figure}[h]
\centering
\includegraphics[width=\hsize]{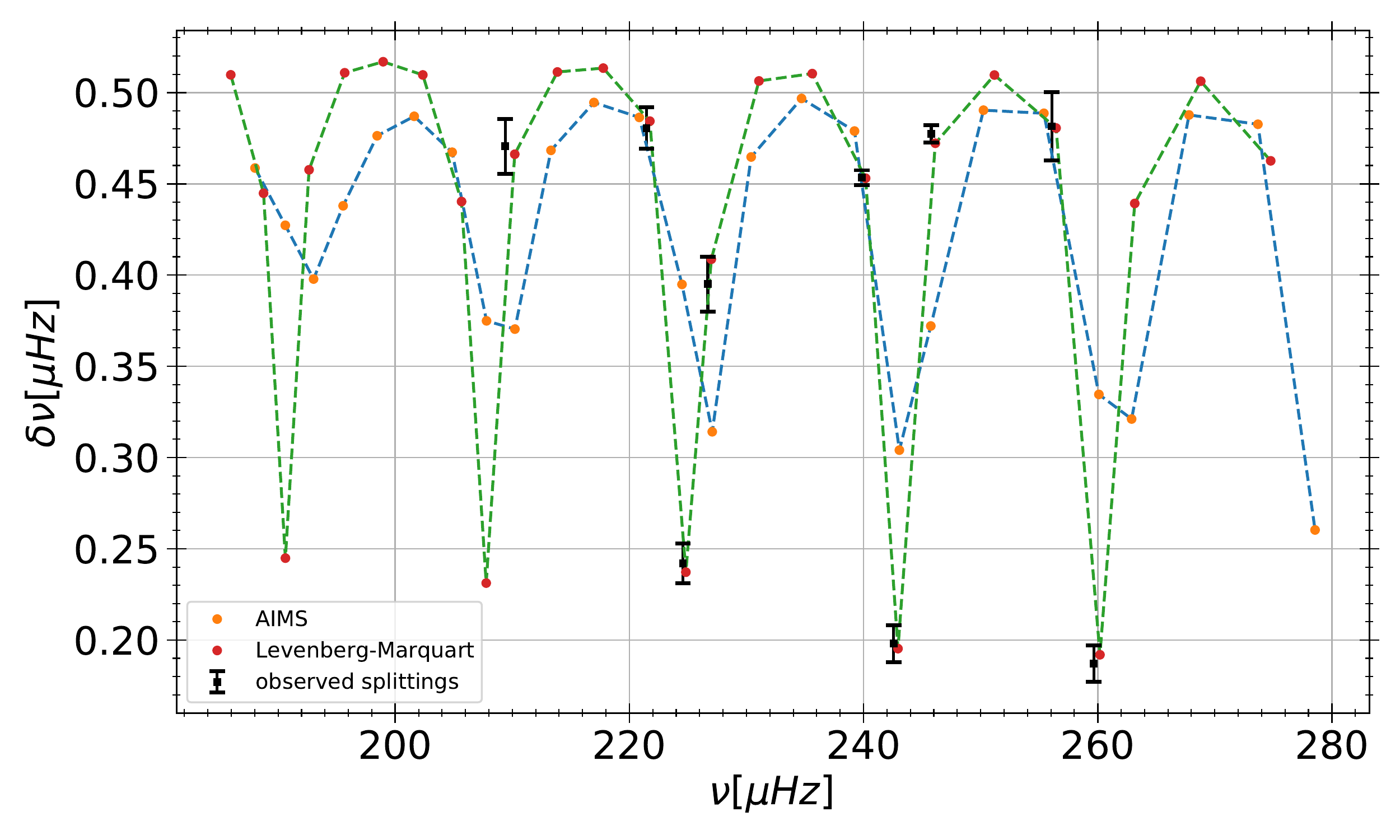}
\caption{Splittings obtained with the two different models presented in Sect. \ref{Stellar_model}, in blue the model obtained with Levenberg-Marquardt minimization technique and in green, the model obtained with AIMS. Splittings were computed, here, with a given rotation profile, thus, the differences seen between the two models can be attributed to the rotational kernel, linked to the structure of the star.}
\label{fig.splitting_spectrum}
\end{figure}

\section{Additional figures}\label{Appendix_figure}

\begin{figure}[h]
\centering
\includegraphics[width=\hsize]{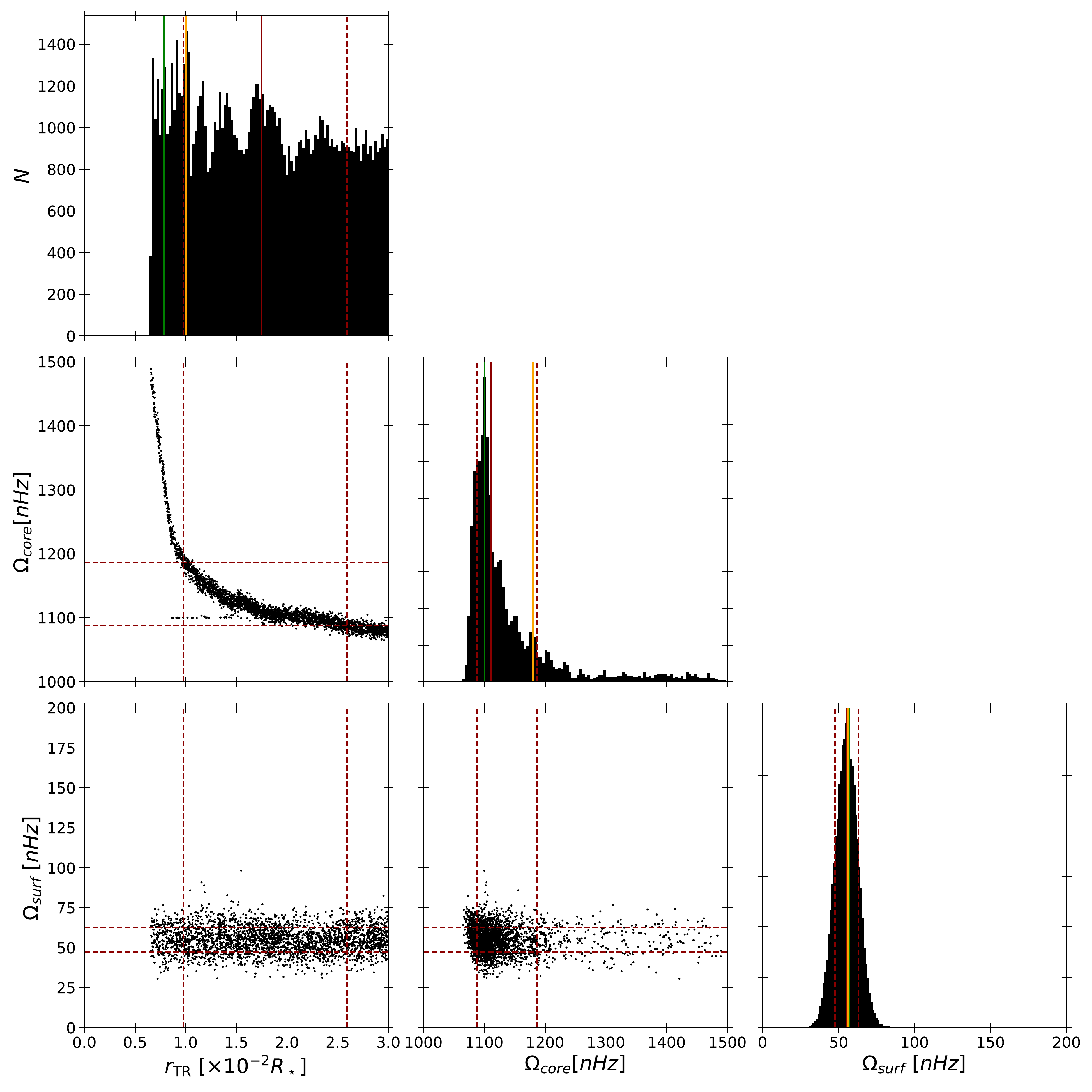}
\caption{Parameter probability distribution obtained with the check-up of MCMC analysis on the step function. The red solid line represents the median of the distribution, the red dotted lines correspond to the $1-\sigma$ uncertainties assuming a Gaussian distribution. The solid green line represent the maximum of the likelihood solution while the solid orange correspond to the initial input parameters that the check-up should recover. Only a small fraction of the sampler is displayed to avoid overweighted figures, however, the histograms were constructed with the full sample.} 
\label{fig._checkup_gate}
\end{figure}

\begin{figure}[h]
\centering
\includegraphics[width=\hsize]{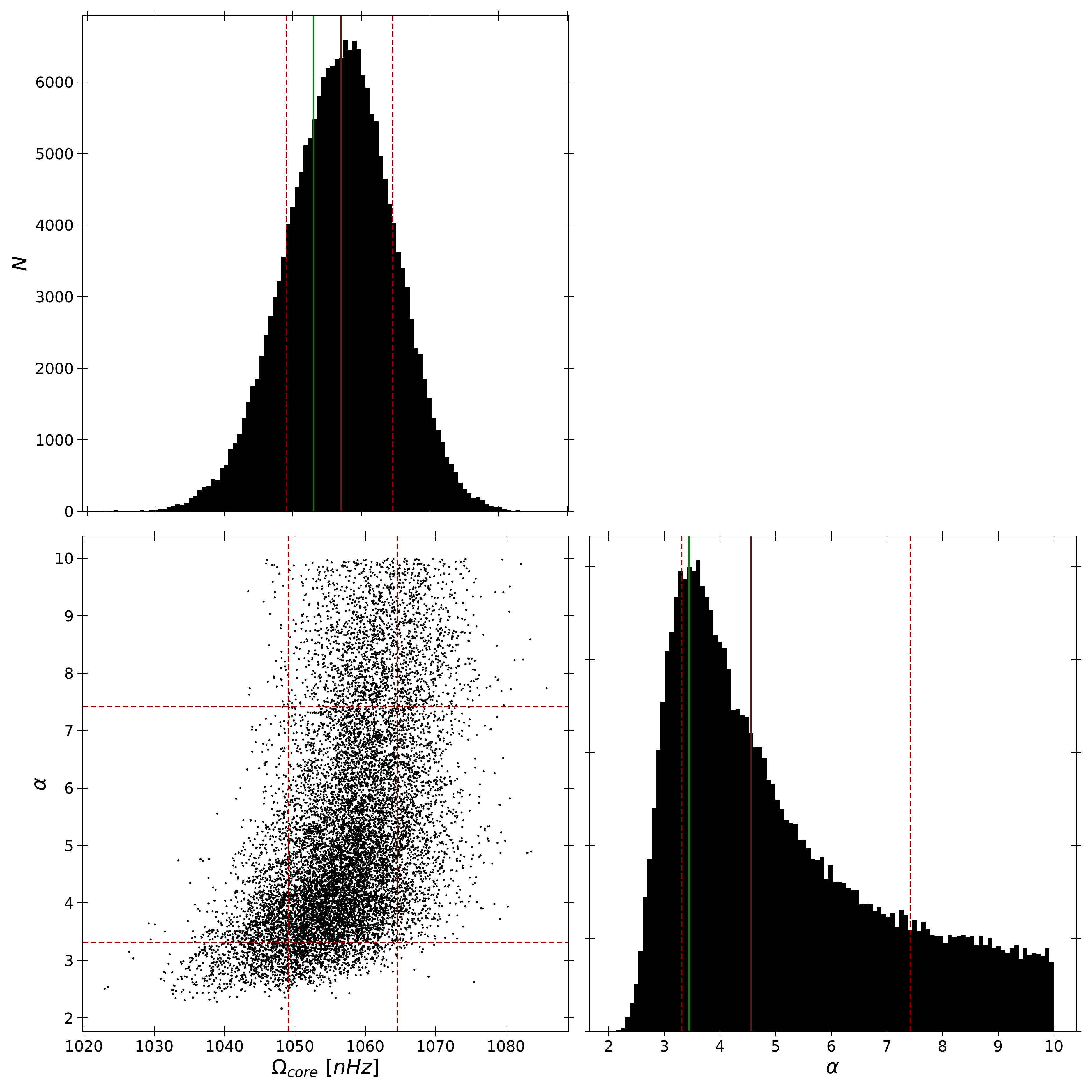}
\caption{Triangle diagram presenting the result from the MCMC analysis on the power law function with the data of Kepler 56. The probability distribution of $\alpha$ is cut at $\alpha=8$ due to a low amplitude noise impacting the computation of the median. The red solid line represents the median, the red dotted lines its one-sigma uncertainties and the green solid line the best MCMC solution.}
\label{fig._MCMC_distri_powerlaw}
\end{figure}

\end{document}